\documentclass[%
reprint,
superscriptaddress,
showpacs,
showkeys,
 amsmath,amssymb,
 aps,
pra,
]{revtex4-1}

\usepackage [latin1]{inputenc}
\usepackage{mathtools}
\usepackage{graphicx}
\usepackage{dcolumn}
\usepackage{bm}
\usepackage{ bbold }
\usepackage{acronym}
\usepackage{color}

\newacro{goe}[GOE]{Gaussian orthogonal ensemble}
\newacro{FOE}[FOE]{fluctuation operator expansion}
\newacro{P}[P]{Poisson}
\newacro{ME}[ME]{mobility edges}
\newacro{MF}[MF]{mean-field}
\newacro{QP}[QP]{quasiparticle}
\newacro{MBL}[MBL]{many-body localization}
\newacro{BG}[BG]{Bose glass}
\newacro{BHM}[BHM]{Bose-Hubbard model}
\newacro{LIOM}[LIOM]{local integrals of motion}
\newacro{SF}[SF]{superfluid}

\begin{document}


\title{Finite-size scaling analysis of localization transitions in the disordered two-dimensional Bose-Hubbard model within the fluctuation operator expansion method}

\author{Andreas Gei{\ss}ler}
\email{andreas.geissler87@gmail.com}
\affiliation{ISIS, University of Strasbourg and CNRS, 67000 Strasbourg, France}
\affiliation{School of Physics \& Astronomy, University of Nottingham, NG7 2RD Nottingham, UK}

\date{\today}
             
\begin{abstract}

The disordered Bose-Hubbard model in two dimensions at non-integer filling admits a superfluid to Bose-glass transition at weak disorder. Less understood are the properties of this system at strong disorder and energy densities corresponding to excited states. In this work we study the Bose-glass transition of the ground state and the related finite energy localization transition, the mobility edge of the quasiparticle spectrum, a critical energy separating extended from localized quasiparticle excitations. To study these the fluctuation operator expansion is used. The level spacing statistics of the quasiparticle excitations, the fractal dimension and decay of the corresponding wave functions are consistent with a many-body mobility edge. The finite-size scaling of the lowest gaps yields a correction to the mean-field prediction of the superfluid to Bose-glass transition. In its vicinity we discuss spectral properties of the ground state in terms of the dynamic structure factor and the spectral function which also shows distinct behavior above and below the mobility edge.
 
\end{abstract}

             
\maketitle


\section{Introduction}

The inclusion of local disorder in the Bose-Hubbard model is able to induce a superfluid to insulator transition at arbitrary filling and low energy densities. The resulting \ac{BG} phase is distinct from the Mott phase at integer filling in that it is nonconducting but has a vanishing gap similar to the \ac{SF} \cite{Hertz1979,Gold1983,Fisher1988,Fisher1989,Bloch2008}. Numerous works have given numerical evidence \cite{Scalettar1991,Buonsante2007,Bissbort2009,Pollet2009,Gurarie2009,Soyler2011} and analytical results \cite{Fisher1988,Fisher1989,Herbut1997,Herbut1998,Lugan2007,Falco2009a,Falco2009,
Ristivojevic2014,Wang2016} showing its existence. In addition the \ac{BG} phase has been probed experimentally in one \cite{Fallani2007} and three dimensional \cite{Meldgin2016} cold atom setups with an optical lattice as well as for bosonic quasiparticles in a doped quantum magnet \cite{Yu2012}. In two dimensions the scaling properties at criticality have been studied extensively in the Bose-Hubbard model \cite{Kisker1997} and its hard-core boson limit \cite{Makivic1993,Zhang1995,Priyadarshee2006,AlvarezZuniga2015} using a wide range of advanced numerical tools, with results comparing mostly quite well with earlier analytical predictions \cite{Fisher1988,Fisher1989,Herbut1997}.

In a recent work we have studied the related localization of \ac{QP} excitations finding that disorder induces \ac{ME} for all values of the local interaction in the full \ac{QP} spectrum of a disordered two dimensional \ac{BHM} \cite{Geissler2019a}. Earlier works have discussed localized \ac{QP}s in one dimensional weakly interacting \ac{BHM}s with correlated speckle \cite{Lugan2007a,Lugan2011} or quasiperiodic potentials \cite{Lellouch2014} involving a \ac{ME} and  with delta-correlated disorder \cite{Bilas2006} but no \ac{ME}. These indicate separate categories of localization as represented by distinct limiting bounds for the correlation length scaling exponent $\nu$ in the respective cases. For delta-correlated disorder the Harris-Chayes-Chayes-Fisher-Spencer bound $\nu > 2/d$ applies
\cite{Harris1974,Chayes1986} whereas for quasiperiodic (thus correlated) disorder the Harris-Luck bound predicts $\nu > 1/d$ \cite{Luck1993} suggesting separate universality classes.
Thus, QP MEs in one dimensional BHMs with quasiperiodic (correlated) disorder do not necessarily imply QP MEs in two dimensional BHMs with delta-correlated disorder \cite{Geissler2019a}. These two universality classes are
also expected to generalize to strong interactions \cite{Khemani2017}, the regime of so-called \ac{MBL} \cite{Altshuler1997,Basko2006,Oganesyan2007,Pal2010}, which has received increased interest in recent years with exciting connections to the fields of topological states \cite{Huse2013,Bauer2013,Decker2019} or quantum computing \cite{Smith2016} to name a few \cite{Nandkishore2015,Abanin2017}. One of its most renown features is its incompatibility with the eigenstate thermalization hypothesis resulting from an extensive number of \ac{LIOM} \cite{Serbyn2013a,Huse2014,Chandran2015a}. A complete demonstration of \ac{MBL} in principle requires complete knowledge of the spectrum, limiting exact diagonalization based analyses to small system sizes \cite{Sierant2018,Wahl2019,Yao2020}. Numerous perturbative arguments \cite{Fleishman1980,Altshuler1997,Basko2006,Nandkishore2014} and increasing numerical evidence \cite{Oganesyan2007,Pal2010,Kshetrimayum2019} have supported its existence in two dimensions, involving a \ac{ME} separating mobile from localized states in the spectrum. Due to its unconstrained local basis, bosonic lattice systems have turned out to be especially hard for numerical simulations, limiting most works to small scale one-dimensional \cite{Sierant2018,Orell2019} and two-dimensional systems with a constrained local basis \cite{Wahl2019}, though strong arguments have been put forward in favor of an \ac{MBL} transition in a disordered continuum system of ultracold bosons in two spatial dimensions \cite{Bertoli2018,Bertoli2019}, even as a function of temperature consistent with a \ac{ME}.

Nevertheless, despite a rigorous proof for certain one-dimensional spin-chains \cite{Imbrie2016,Imbrie2016a}, recent numerical works have challenged the possibility of a thermal phase transition for two dimensional systems \cite{DeRoeck2016,Agarwal2017,DeRoeck2017} and even argued for the absence of a proper localization-delocalization transition in the thermodynamic limit for a one-dimensional spin-chain \cite{Suntajs2019} sparking some counter arguments in \cite{Abanin2019}. Also, it has been argued recently that the neccessary length- and timescales that have to be reached to uniquely identify a \ac{MBL}-type transition are currently out of reach both experimentally and theoretically \cite{Panda2019}. Nevertheless, some experimental realizations have already shown strong signs of localization in cold atom setups, where a disorder potential can be imprinted onto the optical lattice in one \cite{Schreiber2015,Rispoli2018,Lukin2019}, two \cite{Choi2016,Rubio-Abadal2019} and three dimensions \cite{Kondov2015}, as well as for trapped ion \cite{Smith2016} and solid-state spin chains \cite{Wei2018}.

\subsection{System}\label{sec:d-BHM}

In this work we analyze the quantum phases of the disordered two-dimensional \ac{BHM} in order to determine the critical scaling of its ground state \ac{SF} to \ac{BG} transition on a \ac{MF} level and  within the \ac{FOE} method. While we have already discussed the critical scaling at the \ac{ME} in \cite{Geissler2019a}, we focus on low energy excitations to study the localization of many-body \ac{QP} excitations and quantum glass phenomenology on equal footing \cite{Pekker2014,Rademaker2019}.
In second quantization the grand canonical \ac{BHM} with disorder using $\hbar = 1$ can be written as

\begin{align} \label{eq:MBL-hamil}
\hat{H} =  \sum_{\ell}^{L^2} \underbrace{ \left( \mu_{\ell} \hat{b}^{\dag}_{\ell} \hat{b}_{\ell} + \frac{U}{2} \hat{b}^{\dag}_{\ell} \hat{b}^{\dag}_{\ell} \hat{b}_{\ell} \hat{b}_{\ell} \right) }_{\hat{H}_{\ell}} - t \sum_{\langle \ell,\ell' \rangle} (\hat{b}^{\dag}_{\ell} \hat{b}_{\ell'} + \textrm{h.c.} )
\end{align}
with $\mu_{\ell} = -\mu + \epsilon_{\ell}$ given by the local potential $\mu$ and the random potential $\epsilon_{\ell}$, while $U$ and $t$ are the Bose-Hubbard interaction and tunneling rate, respectively. We always choose $\mu$ such that the mean occupation number $n = \langle n_{\ell} \rangle_d = 0.5$ where $\langle \cdot \rangle_d$ is the disorder average and $n_{\ell} = \langle \hat{b}^{\dagger}_{\ell} \hat{b}_{\ell} \rangle$. For $\epsilon_{\ell}$ we assume a Gaussian distribution $P(\epsilon_{\ell}) = \left( 2 \pi W^2 \right)^{-1/2} \exp\left( - \frac{\epsilon_{\ell}^2}{2 W^2} \right)$ as has been realized in recent experiments \cite{Choi2016,Rubio-Abadal2019} with $W$ its standard deviation. This describes a homogeneous system insofar as $\langle \mu_{\ell} \rangle_d = -\mu$. We furthermore consider a simple $L \times L$ square lattice with spacing $a$ and periodic boundary conditions.

The ground state of~\eqref{eq:MBL-hamil} has been investigated in the hard-core limit $U \rightarrow \infty$ to study the \ac{SF} to \ac{BG} transition \cite{Makivic1993,Zhang1995,Priyadarshee2006,AlvarezZuniga2015}. Regarding the regime of moderate interaction strength, it has been shown that due to disorder there is no direct \ac{SF} to Mott insulator phase transition at unit filling \cite{Pollet2009,Gurarie2009,Soyler2011}, which in the ground state instead happens via an intermediate \ac{BG} phase. For non-integer filling or small $U$ there is only the \ac{SF} to \ac{BG} transition.

Here, we evaluate mean-field and quasiparticle spectral properties of the disordered \ac{BHM}~\eqref{eq:MBL-hamil} in terms of the \ac{FOE} method \cite{Bissbort2014,Frerot2016,Geissler2018,Geissler2019a}, a beyond mean-field quasiparticle expansion method. For all disorder strengths we find a critical point in the ground state at sufficiently strong disorder that is consistent with a \ac{SF} to \ac{BG} transition. Considering the fractal dimension of an inhomogeneous Gutzwiller-type mean-field representation of the ground state wave function \cite{Fisher1989,Rokhsar1991,Krauth1992} we find finite-size scaling exponents that match surprisingly well with earlier (analytical) predictions \cite{Fisher1988,Fisher1989,Herbut1997} in contrast to results from more advanced numerical simulations \cite{Priyadarshee2006,AlvarezZuniga2015}. As the FOE method gives access to the complete spectrum of \ac{QP}s, we use it to discuss spectral properties of experimental interest by considering the beyond mean-field \ac{QP} ground state. We note that all \ac{QP} excitations tend to resemble approximate \ac{LIOM} for sufficiently strong disorder.

\subsection{Overview}

The remainder of this work is structured in four main sections and a summary. First, we determine the mean-field ground state of the disordered BHM in order to characterize the \ac{SF} to \ac{BG} transition in terms of the Edwards-Anderson parameter and the fractal dimension in Sec.~\ref{sec:d-BHM_MF_BG}. In particular we determine the finite-size scaling collapse for the fractal dimension of the \ac{MF} ground state condensate order parameter. Next, we detail the \ac{FOE} method used for the remainder of this work to determine the quasiparticle spectrum beyond the weak-coupling ansatz of the Bogoliubov method. In Sec.~\ref{sec:Commutator_deviation} and Sec.\ref{sec:QP_diagonalization} we also discuss numerical tests of its applicability for the disordered BHM~\eqref{eq:MBL-hamil}. The following Sec.~\ref{sec:FOE-charact_qp_spec} focuses on a detailed discussion of the full quasiparticle spectrum. There, we discuss the energy level statistics and localization properties of the fluctuation wave functions in order to discern localized and non-local states separated by a \ac{ME}. By considering a simple finite-size scaling ansatz we further establish a relation between the lowest excited \ac{QP} fluctuation states and the \ac{SF} to \ac{BG} transition in the ground state. In the final Sec.~\ref{sec:spec_func} we consider the spectral properties of the \ac{FOE}'s quasiparticle ground state in the vicinity of the \ac{SF} to \ac{BG} phase transition which nicely reflect the phenomenology discussed in the previous sections. We end with a brief summary in Sec.~\ref{sec:summary}.

\section{Mean-field critical point} \label{sec:d-BHM_MF_BG}

We start by characterizing the ground state properties of~\eqref{eq:MBL-hamil}, specifically in relation to the aforementioned occurrence of a Bose-glass phase \cite{Kisker1997,Makivic1993,Zhang1995,Priyadarshee2006,AlvarezZuniga2015} in and close to the ground state. Here, we consider a simple Gutzwiller \ac{MF} product ansatz of the form $| \psi_{\textrm{MF}} \rangle = \prod_{\ell} | \psi_0 \rangle_{\ell}$ where each $| \psi_0 \rangle_{\ell}$ is given in terms of a linear combination over the local Fock-basis truncated at some fixed number $N_b$. Throughout this work at least a value of $N_b=9$ or greater is used, sufficient to guarantee convergence of the mean-field ground state and the lowest local Gutzwiller excitations discussed in Sec.~\ref{sec:FOE_Gops}. Their, in general, complex amplitudes can either be found via a minimization of the energy or a self-consistent procedure (see Sec.~\ref{sec:FOE_Gops}). On this mean-field level we focus on two observables to characterize the occurrence of a transition point in the ground state phase for an increasing disorder potential, where a \ac{SF} to \ac{BG} transition is expected. We note that the superfluid fraction is expected to vanish at this transition while the condensate fraction is not. While the former can be determined using twisted boundary conditions \cite{Buonsante2007}, here we consider complementary observables which are more closely related to previous works \cite{Geissler2019a}. Firstly, we define an Edwards-Anderson-type order parameter 
\begin{align}
q_{\textrm{EA}} = \frac{1}{L^2} \sum_{\ell} \left( \langle n_{\ell} n_{\ell} \rangle_d - \langle n_{\ell} \rangle_d \langle n_{\ell} \rangle_d \right)
\end{align}
with $n_{\ell} = \langle \hat{n}_{\ell} \rangle$ the expectation value of the local boson number density and $\langle \cdot \rangle_d$ the disorder average. By construction it is always zero in a homogeneous state and only non-zero if the correlations between the density and the disorder are extensive \cite{Morrison2008,Thomson2016}. Furthermore, we consider the fractal dimension $D_{\phi}$ \cite{Castellani1986,Serbyn2016} of the condensate wave function $\phi_{\ell} = \langle \hat{b}_{\ell} \rangle$, for which we use the definition \cite{Mace2018,Lindinger2019}

\begin{align}
D_{\phi} = \left\langle \log_L \left( \frac{ \sum_{\ell}^{L^2} |\mathbf{\phi}_{\ell}|^2}{\textrm{max}_{\ell}|\mathbf{\phi}_{\ell}|^2 } \right) \right\rangle_d.
\end{align}
We evaluate both characteristics over a range of parameters $U/t \in [1,25]$ and $W/t \in [0,15]$, and for the linear system sizes $L \in \mathcal{L} = \lbrace 10,20,24,32,40 \rbrace$ while averaging over $N_r = 60$ disorder realizations each time.

\begin{figure}[t]
 \centering
 \includegraphics[width=0.7\columnwidth]{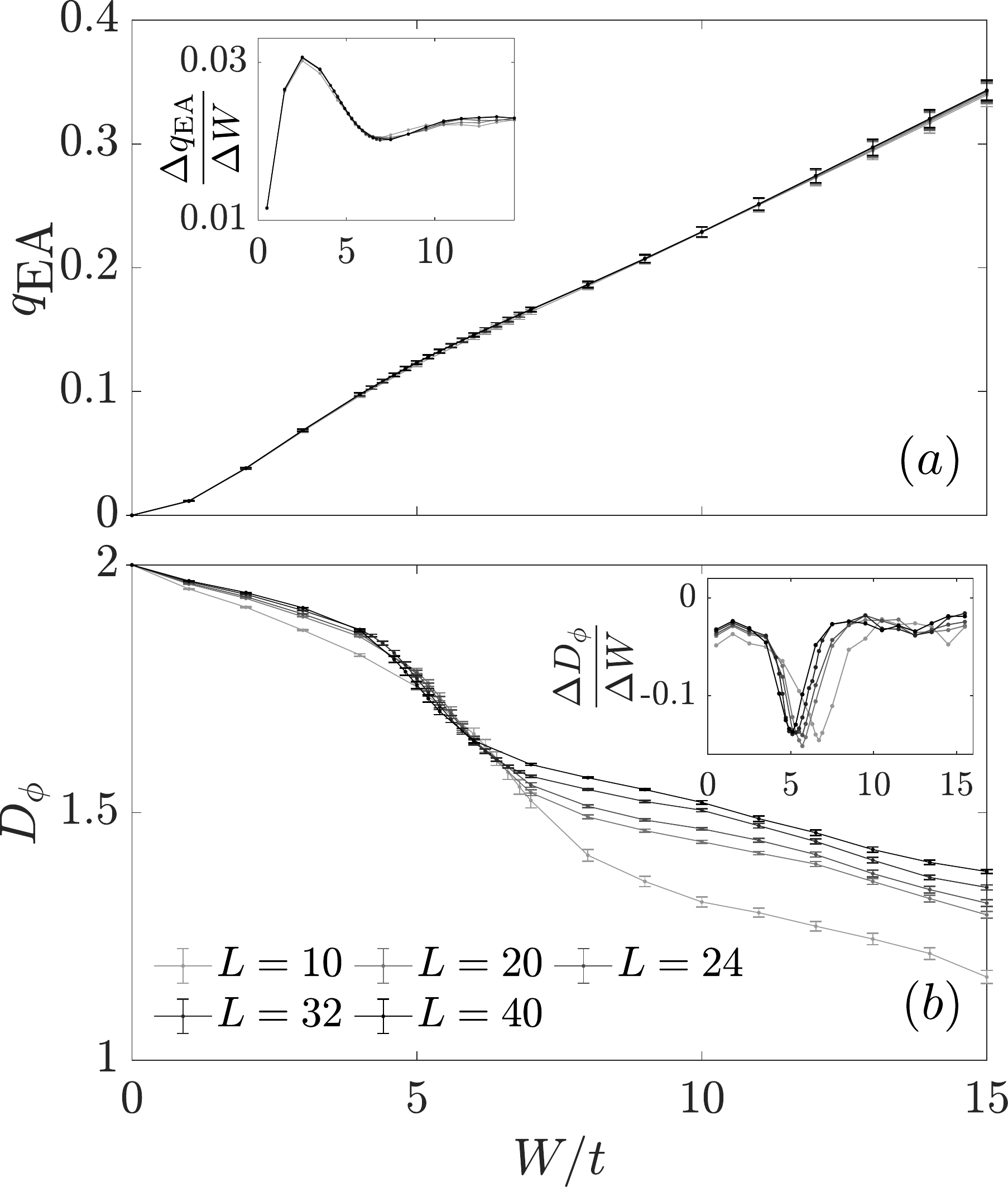}
 \caption{
Characterization of the \ac{MF} critical point for the 2D BHM with disorder. The Edwards-Anderson parameter $q_{\textrm{EA}}$ $(a)$ and the fractal dimension $D_{\phi}$ $(b)$ are shown together with their respective numerical derivatives in the insets. Both are given as a function of the disorder $W$ for fixed interaction $U/t = 20$ and various system sizes (see legend).
 }
 \label{fig:MF_GS_obs}
\end{figure}

As an example we show $q_{\textrm{EA}}$ and $D_{\phi}$ for $U/t = 20$ in Fig.~\ref{fig:MF_GS_obs}. As discussed in \cite{Morrison2008} $q_{\textrm{EA}}$ in panel $(a)$ is non-zero for all $W>0$. While this parameter is almost independent from the considered system sizes, it also barely exhibits any extremal behavior except for the soft kink at $W/t \approx 5$ visible in the numerical derivative $\Delta q_{\textrm{EA}} / \Delta W$ [inset Fig.~\ref{fig:MF_GS_obs}$(a)$]. Still, a nonzero value of $q_{\textrm{EA}}$ indicates the occurrence a glassy ground state for increasing disorder. The fractal dimension $D_{\phi}$ in panel $(b)$, on the other hand, features a much more pronounced drop in the same disorder range, suggesting the presence of a phase transition, usually accompanied by finite-size scaling effects in the vicinity of the critical point. 
In order to quantify this scaling we first consider the numerical derivative $\Delta D_{\phi} / \Delta W$ [see inset Fig.~\ref{fig:MF_GS_obs}$(b)$], which exhibits a minimum corresponding to an inflection point of $D_{\phi}(W)$ at $W_0$ that shifts to small disorder strength for increasing system sizes resulting in a finite-size scaling of $D_0(L) \equiv D_{\phi}\left[W_0(L)\right]$ [see Fig.~\ref{fig:MF_GS_scaling}$(a)$]. We observe such a minimum for all $U/t \gtrsim 10$. 

\subsection{Finite-size scaling}

Such a finite-size shift indicates a critical point with a scaling that is typically of the form \cite{AlvarezZuniga2015}
\begin{align} \label{eq:BG-collapse}
D_{\phi;L,U}(W) - D_c = L^{-\alpha} \tilde{D}_U \left( \left[ W-W_c(U) \right] L^{1/\nu} \right),
\end{align}
with the critical fractal dimension $D_c$, the critical disorder $W_c(U)$, a universal function $\tilde{D}_U$ with parameter $U$, as well as the critical exponents $\alpha$ and $\nu$.
For the scaling collapse of the inflection points $D_{0;U}(L)$ onto the inflection point of the scaling function $\tilde{D}_U(\bar{W}_0)$, where $\bar{W}_0 = \left[W_0(L)-W_c\right] L^{1/\nu}$ is the rescaled disorder, we thus expect
\begin{align}
D_{0;U}(L) = \frac{\tilde{D}_U(\bar{W}_0)}{L^{\alpha}} + D_c. \label{eq:collapse_Dphi_inflection}
\end{align}
As this expression has three unknown parameters, compared to the five system sizes $L \in \mathcal{L} = \lbrace 10,20,24,32,40 \rbrace$ considered for each value of $U$, we first determine the best fit parameters $\tilde{D}_U(\bar{W}_0)$ and $D_c$ for fixed values of $\alpha$ and $U/t \in \lbrace 15,20,25 \rbrace$ to obtain the functional relation $D_c(\alpha)$ shown in Fig.~\ref{fig:MF_GS_scaling}$(b)$, while exemplary fits for $\alpha = 0.44$ are shown in Fig.~\ref{fig:MF_GS_scaling}$(a)$. By definition $D_c$ is limited from above so the collapse of the inflection points gives a lower bound $\alpha > 0.4$ [see Fig.~\ref{fig:MF_GS_scaling}($b$)]. As the finite-size scaling Eq.~\eqref{eq:BG-collapse} is independent of the scaling exponent $\nu$ at the critical point $W_c(U)$, we can further determine $W_c(U)$ if we scale only the fractal dimension according to $(D_{\phi}-D_c) L^{\alpha}$ to obtain the crossing point of all system sizes, as shown in the inset of Fig.~\ref{fig:MF_GS_scaling}($c$). This way we get the best candidates for the critical point $(W_c(U),D_c)$ as a function of $\alpha$, exemplary depicted in Fig.~\ref{fig:MF_GS_scaling}$(b)$ for $U/t = 20$. To quantify the goodness of these fits we consider the adjusted coefficient of determination $\tilde{R}^2$ given in the inset of Fig.~\ref{fig:MF_GS_scaling}$(b)$ with errorbars representing the standard deviation when sampling over $U/t \in \lbrace 15,20,25 \rbrace$ and six distinct subsets of 10 disorder realizations each. The value of $\tilde{R}^2$ for these fits is almost constantly at its optimum for the considered range of $\alpha$.

For the full collapse we only have to consider $\alpha$ and $\nu$ in order to minimize the mean relative variance as a measure for the goodness of the collapse:
\begin{widetext}
\begin{align} 
\chi_{D_{\phi}} = \sum_{U} \frac{\chi_{D_{\phi}}(U)}{N_U} = \sum_{U} \sum_{L' > L} \sum_{\bar{W}} \frac{\left({D}_{\phi;L,U} (\bar{W})  - {D}_{\phi;L',U} (\bar{W})\right)^2}{2 \left[ \sigma{D}^2_{\phi;L,U} (\bar{W}) + \sigma{D}^2_{\phi;L',U} (\bar{W})\right]} \frac{1}{\bar{C}_{D_\phi}}. \label{eq:alg_Dphi_mean_rel_dev} 
\end{align}
\end{widetext}
Here, $\sigma{D}_{\phi;L,U} (\bar{W})$ are the standard errors of the mean determined from the disorder sampling while the normalization constant $\bar{C}_{D_\phi} $ is given by the total number of terms, $\bar{C}_{D_\phi} = N_U \sum_{L' > L} \sum_{\bar{W}} 1$ with $N_U$ the number of considered interaction values. For an ideal collapse this measure should be on the order of $1$. In order to estimate the error of the obtained scaling exponents this finite-size scaling procedure is repeated for 6 independent subsets of 10 disorder realizations each, while the interaction sum takes into account all considered values $U/t \in \lbrace 1,3,5,10,15,20,25 \rbrace$. The free parameters of this collapse are $\nu$ and $\alpha$, the latter of which implicitly determines $D_c(\alpha)$ via the scaling of the inflection points [see Figs.~\ref{fig:MF_GS_scaling}$(a,b)$] as well as $W_c(U)$ via the unique crossing point of the rescaled fractal dimension [as in inset Fig.~\ref{fig:MF_GS_scaling}$(c)$].

\subsection{Results}

\begin{figure}[h]
 \centering
 \includegraphics[width=0.99\columnwidth]{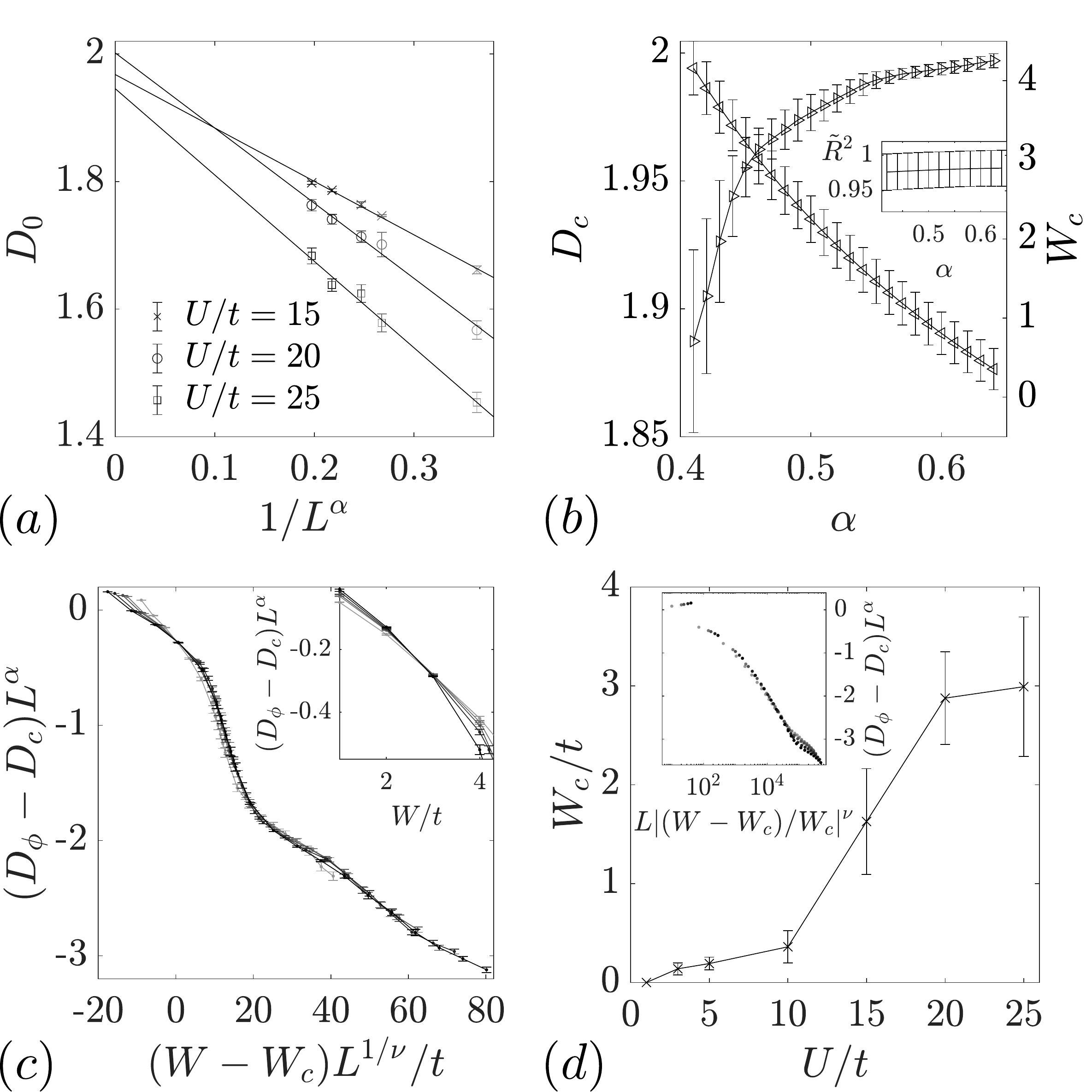}
 \caption{
Finite-size scaling and critical points for the \ac{MF} ground state of the disordered 2D \ac{BHM}. $(a)$ shows the shift of $D_0$ at the inflection point as function of $L$ for $\alpha = 0.44$. Grey colors correspond to $L$ as in Fig.~\ref{fig:MF_GS_obs}$(b)$ and lines are best fits to Eq.~\eqref{eq:collapse_Dphi_inflection} used to determine $D_0$ in the limit $L \rightarrow \infty$. Corresponding best infinite system size predictions for $D_c$ (left pointing triangles) and $W_c(U)$ (right pointing triangles) as determined by the crossing point in the partial collapse [inset $(c)$] is shown in $(b)$ for fixed values of $\alpha$ and $U/t=20$. The adjusted parameter of convergence of these fits is given in the inset of $(b)$. $(c)$ shows the scaling collapse~\eqref{eq:BG-collapse} of $D_{\phi}$ as a function of the rescaled disorder (unscaled in the inset) for $W_c/t = 2.9$ and $U/t = 20$. $(d)$ depicts the \ac{MF} critical disorder $W_c$ determined via the collapse of $D_{\phi}$ according to Eq.~\eqref{eq:BG-collapse} and a corresponding collapse for $U/t = 3$ in the inset.
 }
 \label{fig:MF_GS_scaling}
\end{figure}

An exemplary collapse for $U/t=20$ is given in Fig.~\ref{fig:MF_GS_scaling}$(c)$ which has the individual relative variance $\chi_{D_{\phi}}(U=20t) = 0.44(15)$. In combination the mean relative variance Eq.~\eqref{eq:alg_Dphi_mean_rel_dev} for all interaction values together is $\chi_{D_{\phi}} = 2.8(5)$. It is greater then one primarily due to substantial finite-size corrections far from the critical point at weak interaction resulting in $\chi_{D_{\phi}}(U=3t) = 11(2)$ [inset of Fig.~\ref{fig:MF_GS_scaling}$(d)$]. For all best collapses taken together we find the scaling exponents
\begin{align}
\alpha &= 0.44(2), & \nu &= 2.0(2)
\end{align}
and a critical fractal dimension $D_c^{\phi}/t = 1.97(3)$ indistinguishable from its upper limit. The corresponding critical line $W_c(U)$ is depicted in Fig.~\ref{fig:MF_GS_scaling}$(d)$. 

To summarize, for weak interaction $U/t \lesssim 10$ the critical disorder strength is close to zero. At strong interaction values $U/t \gtrsim 20$, on the other hand, we find a ground state transition point that is consistent with previous predictions of a superfluid to Bose-glass transition also at half-filling but in the hard-core boson limit $U/t \rightarrow \infty$ with box-disorder $\epsilon_{\ell} \in [-W,W]$ for the local potential  \cite{Makivic1993,Zhang1995,Priyadarshee2006,AlvarezZuniga2015}. Additionally, considering earlier results for this system \cite{Fisher1988,Fisher1989,Herbut1997} and the nonzero $q_{\textrm{EA}}$ for $W > W_c(U)$ we associate this critical line with a \ac{SF} to \ac{BG} transition. Notably, the \ac{MF} scaling exponents we find match some early Monte-Carlo predictions surprisingly well \cite{Makivic1993}.

\section{Fluctuation operator expansion}\label{sec:FOE}

We now discuss the fluctuation operator expansion (FOE) \cite{Bissbort2014,Geissler2018} with a main focus on its application to systems with broken translational invariance such as~\eqref{eq:MBL-hamil}, in order to investigate its properties beyond the previous discussion of the \ac{MF} ground state. Given any such \ac{MF} state the \ac{FOE} constitutes a systematic expansion of all beyond first-order fluctuation operator terms of ~\eqref{eq:MBL-hamil} -- commonly consisting of only non-local terms neglected on the \ac{MF} level -- in terms of a quadratic map onto local complete sets of generators of \ac{MF} excitations, the Gutzwiller operators. Within the approximation of a negligible small density of local Gutzwiller fluctuations these operators are quasi-bosonic and their second-order contribution to original Hamiltonian -- beyond the mean-field terms -- can be brought into a diagonalizable Nambu-type form. Its diagonalization results in pairs of \ac{QP} excitation energies $\omega_{\gamma}$ and $-\omega_{\gamma}^*$ with corresponding wavefunctions $\mathbf{x}^{(\gamma)}$ and $\mathbf{y}^{(\gamma)}$, which allow for a characterization of the spectrum \footnote{It is furthermore straightforward to determine any type of correlation given in terms of non-local products of local operators.}.

\subsection{Gutzwiller operator representation}\label{sec:FOE_Gops}
 
The FOE is a quasiparticle method based on an expansion of a second quantized Hamiltonian such as~\eqref{eq:MBL-hamil} in terms of the eigenstates $|i \rangle_{\ell}$ of its local mean-field Hamiltonians (given a truncation $N_b$ of the local bosonic number states) 
\begin{align}
\hat{H}_{\textrm{MF}}^{(\ell)} = \hat{H}_{\ell} - t \sum_{ \lbrace \ell' | \langle \ell, \ell' \rangle \rbrace} \left( \hat{b}^{\dag}_{\ell} \phi_{\ell'} + \textrm{h.c.} \right). \label{eq:MF-Hamil}
\end{align}
These are defined in terms of the fluctuation operators  $\hat{\delta b}_{\ell} \equiv \hat{b}_{\ell} - \phi_{\ell}$ and the complex fields $\phi_{\ell} \stackrel{!}{=} {}_{\ell}\langle 0| \hat{b}_{\ell} |0 \rangle_{\ell}$ which are to be determined self-consistently \footnote{We note that self-consistency is a neccessary but not sufficient condition when determining the mean-field ground state. This implies that the FOE can just as well be used to characterize fluctuations of mean-field-type states far from the ground state.}. Drawing from variational concepts \cite{Huber2007,Bissbort2011,Endres2012} the FOE allows for a systematic improvement over standard Bogoliubov theory \cite{Bogolyubov1947} by considering in principle general local fluctuations, giving access to the complete QP spectrum of the original Hamiltonian~\eqref{eq:MBL-hamil}
\begin{align}
\hat{H} = \sum_{\ell} \hat{H}_{\textrm{MF}}^{(\ell)} - t \sum_{\langle \ell, \ell' \rangle} \left( \hat{\delta b}_{\ell}^{\dag} \hat{\delta b}_{\ell'} - \phi_{\ell}^* \phi_{\ell'} + \textrm{h.c.} \right).
\end{align}

Due to the completeness of each local eigenbasis $\lbrace |i \rangle_{\ell} \rbrace$ with eigenenergies $E^{(\ell)}_i$, the FOE representation $\hat{\delta b}_{\ell} = \sum_{i,j<N} {}_{\ell}\langle i | \hat{\delta b}_{\ell} | j \rangle_{\ell} |i \rangle_{\ell}  {}_{\ell}\langle j|$ constitutes a \textit{quadratic} map that is exact in the limit $N \rightarrow \infty$ with $N$ the truncation of the local Gutzwiller eigenbases. To ensure convergence of these bases $N_b = 3 N$ is usually sufficient. It is convenient to introduce the local Gutzwiller raising and lowering operators as well as their compound terms for all $i>0$:

\begin{align}
\sigma_{\ell}^{(i)^{\dagger}} &\equiv |i \rangle_{\ell}  {}_{\ell}\langle 0|, &
\mathbb{1}_N - \sum_{i>0}^N \sigma_{\ell}^{(i)^{\dagger}} \sigma_{\ell}^{(i)} &= |0 \rangle_{\ell}  {}_{\ell}\langle 0|, \\
\sigma_{\ell}^{(i)} &\equiv |0 \rangle_{\ell}  {}_{\ell}\langle i|, & \textrm{and} \quad
\sigma_{\ell}^{(i)^{\dagger}} \sigma_{\ell}^{(j)} &= |i \rangle_{\ell}  {}_{\ell}\langle j|.
\end{align}
Using these operators one obtains the formally exact representation $\hat{H} = \sum_{\ell} \hat{H}_{\textrm{MF}}^{(\ell)} + \mathcal{H}^{(2)} + \mathcal{H}^{(3)} + \mathcal{H}^{(4)}$, where each term $\mathcal{H}^{(n)}$ refers to a different order $n$ in the Gutzwiller operators. We note that the self-consistency condition guarantees the absence of first order terms. While the second order term $\mathcal{H}^{(2)}$ yields the full spectrum of non-interacting QP fluctuations, the higher order terms introduce interactions among them. A sufficiently low density of local Gutzwiller excitations implies that the interaction terms can be neglected. We will thus consider the beyond mean-field Hamiltonian $\hat{H}^{(2)} \equiv \sum_{\ell} \hat{H}_{\textrm{MF}}^{(\ell)} + \mathcal{H}^{(2)}$. This can be justified in the vicinity of the ground state that can implicitly be defined as the state not containing any \ac{QP} excitations (discussed in Secs.~\ref{sec:Commutator_deviation} and~\ref{sec:QP_diagonalization}), resulting in very good predictions both in Mott-type and superfluid phases \cite{Bissbort2012,Geissler2018T,Geissler2018}. Furthermore, in the localized phase at strong disorder the eigenstates of this approximate Hamiltonian tend to display similarities to approximate \ac{LIOM}s, as discussed in Sec.~\ref{sec:FOE-charact_qp_spec}. These also have the property that their spectra are (nearly) unaffected by one another, resulting in the absence of level repulsion in the localized regime \cite{Maksymov2019} and causing the well-known Poisson statistics of the level spacings also discussed in Sec.~\ref{sec:FOE-charact_qp_spec}.

\subsection{Quasi-Bosonic commutation relations}\label{sec:Commutator_deviation}

Before we can attempt to diagonalize $\hat{H}^{(2)}$ we first have to bring it into a standard Nambu-type form, which is straightforward for regular bosons. To do so in our case we have to consider the actual commutation relations that characterize the Gutzwiller operators ${\sigma}_{\ell}^{(i)^{\dag}}$ and ${\sigma}_{\ell}^{(j)}$. One can easily show that they obey quasi-bosonic commutation relations of the form 
\begin{align}
 \left[ {\sigma}_{\ell'}^{{(j)}} , {\sigma}_{\ell}^{{(i)}^{\dag}} \right] &= \delta_{i,j} \delta_{\ell,\ell'} - \delta_{\ell,\ell'} \hat{R}^{(i,j)}(\ell), 
\label{eq:GWfluc_commut_red} \\
\left[ {\sigma}_{\ell'}^{{(i)}^{\dag}} , {\sigma}_{\ell}^{{(j)}^{\dag}} \right] &= \left[ {\sigma}_{\ell'}^{{(i)}} , {\sigma}_{\ell}^{{(j)}} \right] = 0.
 \label{eq:GWfluc_commut_zero}
\end{align}
Here, we introduce the residual operator $\hat{R}^{(i,j)}(\ell)$ quantifying the deviation from bosonic behavior. It is given by the expression
\begin{align}
\hat{R}^{(i,j)}(\ell) \equiv& {\sigma}_{\ell}^{{(i)}^{\dag}} {\sigma}_{\ell}^{(j)} + \delta_{i,j} \sum_{j'>0} {\sigma}_{\ell}^{{(j')}^{\dag}} {\sigma}_{\ell}^{(j')}, \label{eq:R_of_l}
\end{align} 
which is on the order of the local occupation of Gutzwiller fluctuations $\kappa_{\ell} = \sum_{i>0} \sigma_{\ell}^{(i)^{\dag}} \sigma_{\ell}^{(i)}$. The essential approximation of the FOE method amounts to taking the limit $\langle \hat{R}^{(i,j)}(\ell) \rangle \rightarrow 0$, following from the assumption of only sparsely populated excited Gutzwiller modes, consistent with neglecting the interactions between the local fluctuations $\mathcal{H}^{(3)}$ and $\mathcal{H}^{(4)}$.

\begin{figure}[t]
 \centering
 \includegraphics[width=0.99\columnwidth]{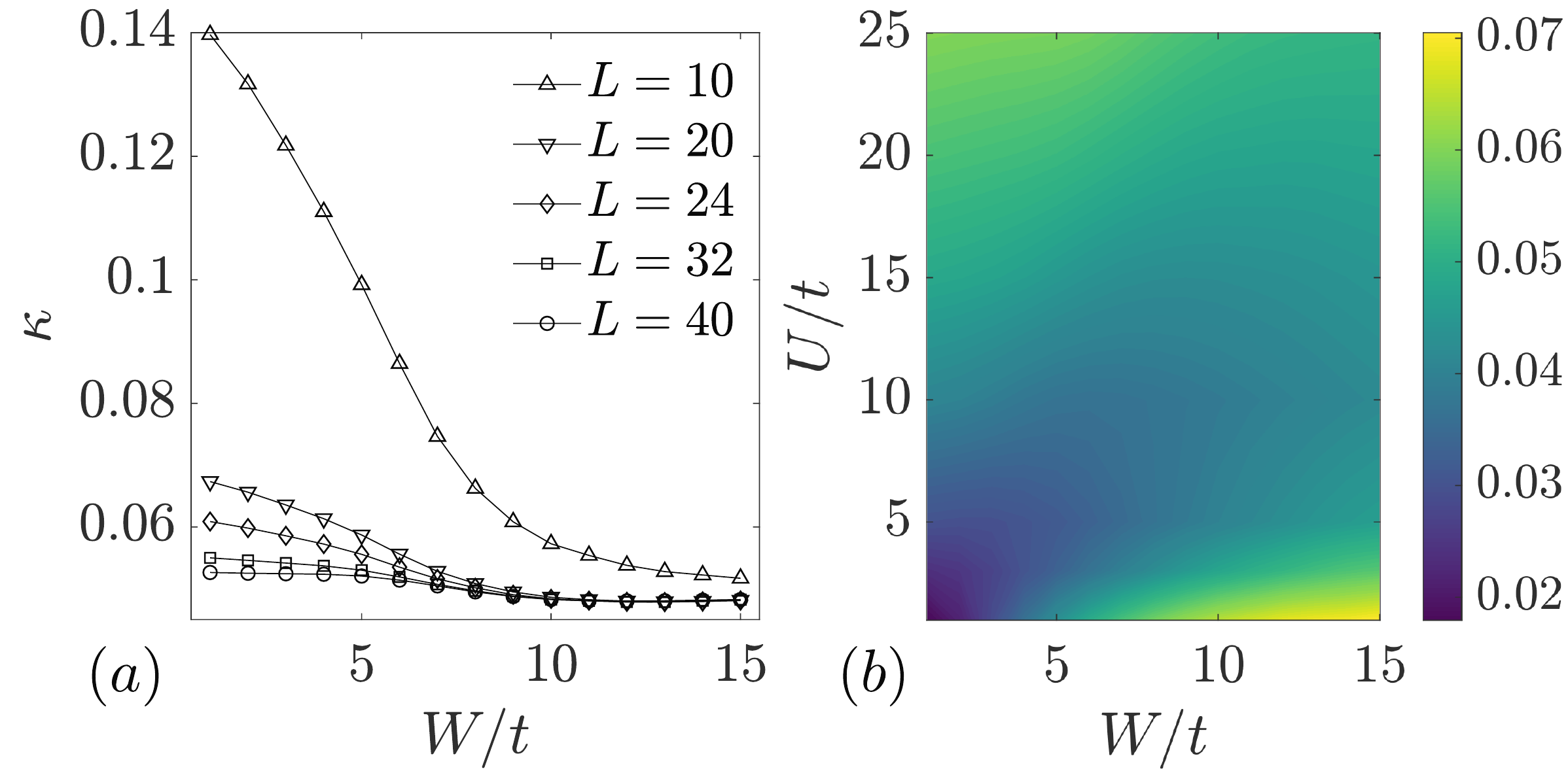}
 \caption{
 Disorder averaged mean fraction of local fluctuations $\kappa$ quantifying the goodness of the FOE. $(a)$ $\kappa$ as a function the disorder $W/t$ for fixed $U/t = 20$, $N=3$ and various system sizes given in the legend. $(b)$ Contour plot of $\kappa$ as function of $W/t$ and $U/t$ for fixed system size $L=32$ and $N=3$.
 }
 \label{fig:kappa}
\end{figure}

We note that \eqref{eq:R_of_l} implies the relation $ \delta_{i,j} \langle \kappa_{\ell} \rangle < \langle \hat{R}^{(i,j)}(\ell) \rangle < (1 +  \delta_{i,j}) \langle \kappa_{\ell} \rangle $, so the approximation can be quantified \textit{a posteriori} via the expectation value $\langle \kappa_{\ell} \rangle$ of the local population of Gutzwiller excitations. It is given in terms of the local overlap of the state in question with the mean-field ground state due to the identity $ \langle \kappa_{\ell} \rangle = \langle \mathbb{1}_N - |0 \rangle_{\ell}  {}_{\ell}\langle 0| \rangle = \sum_{\gamma} \sum_{i>0} |v^{(\gamma)}_{i,\ell}|^2 $. The latter identity is the result for the QP ground state $| \psi_{\textrm{QP}} \rangle$ which we will define in Sec.~\ref{sec:QP_diagonalization}. From this we determine the disorder averaged mean fraction of local fluctuations $\kappa \equiv \sum_{\ell} \langle \langle \kappa_{\ell} \rangle \rangle_d / L^2$, which is shown in Fig.~\ref{fig:kappa} for the complete parameter ranges considered in this work as well as in Ref.\cite{Geissler2019a}. 

For $U/t = 20$, $N=3$, $N_b=3N=9$ as well as half-filling and increasing system sizes $L$ one can see that $\kappa$ strongly decreases down to a limiting value of $\kappa \lesssim 0.05$ [see Fig.~\ref{fig:kappa}$(a)$]. Also for $U/t \in \left[ 1,25 \right] $, $W/t \in \left[ 1,15 \right]$ and fixed $L = 32$ [see Fig.~\ref{fig:kappa}$(b)$] we always find $\kappa \lesssim 0.07 \ll 1$, thus validating the quality of the FOE approximation. With this in mind we can confidently discuss the diagonalization of the quasiparticle Hamiltonian, but first we give a short discussion of its form.

\subsection{The quasiparticle Hamiltonian $\hat{H}^{(2)}$} \label{sec:FOE_qpHamil}

In this section we discuss the second order quasiparticle Hamiltonian. It has a simple bilinear form which can be written in terms of the vectors $ \pmb{\sigma} = \left( \sigma_{1}^{(1)} , \ldots  , \sigma_{L}^{(N-1)} \right)^{\textrm{T}}$ and $ \pmb{\sigma}^{\dag} = \left( \sigma_{1}^{(1)^{\dag}} , \ldots  , \sigma_{L}^{(N-1)^{\dag}} \right)^{\textrm{T}}$. Using these and the approximation $ \langle \hat{R}^{(i,j)}(\ell) \rangle \rightarrow 0$ in the commutation relation~\eqref{eq:GWfluc_commut_red} one can bring the Hamiltonian into a Nambu-type form, so 

\begin{align}
\hat{H}^{(2)} \approx \hat{\mathcal{H}}^{(2)}_{\textrm{QP}} \equiv \frac{1}{2} \begin{pmatrix}
{\pmb{\sigma}} \\ {\pmb{\sigma}}^{\dag}
\end{pmatrix} ^{\dag} &\mathcal{H}^{(2)}_{\textrm{QP}} \begin{pmatrix}
{\pmb{\sigma}} \\ {\pmb{\sigma}}^{\dag}
\end{pmatrix} -\frac{1}{2} \textrm{Tr}(h), \label{eq:H2_in_quasimomentum} \\
\textrm{with} \quad
&\mathcal{H}^{(2)}_{\textrm{QP}} = 
\begin{pmatrix}
{h} & {\Delta} \\
{\Delta}^* & {h}^*  
\end{pmatrix}.
\label{eq:operatorQPH}
\end{align}
As we have to get half of the normal ordered pairs $\pmb{\sigma}^{\dag} \pmb{\sigma}$ into anti-normal order, we obtain the scalar term $\textrm{Tr}(h)/2$ along the way. Within this approximation the introduced Hamiltonian matrix $\mathcal{H}^{(2)}_{\textrm{QP}}$ has a size of $2(N-1)L^2  \times 2(N-1)L^2 $. Its individual entries are given in terms of $\prescript{}{\ell}{\langle} i | \hat{  b}_{\ell} | j \rangle_{\ell}$ matrix elements, each within the local Gutzwiller bases, so the explicit matrix entries are given by

\begin{align}
{h}_{(i,\ell),(j,\ell')} &= - t_{\ell,\ell'} F_{i,0,0,j}^{(\ell,\ell')} + \delta_{\ell,\ell'} \delta_{i,j} E_i^{(\ell)}, \label{eq:QPmatrix_h} \\
{\Delta}_{(i,\ell),(j,\ell')} &= - t_{\ell,\ell'} F_{i,j,0,0}^{(\ell,\ell')}. \label{eq:QPmatrix_Delta}
\end{align}
Both expressions are given in terms of the tunneling matrix, whose matrix elements $t_{\ell,\ell'} = t \, \forall \, \{(\ell,\ell') | \langle \ell,\ell' \rangle\}$ are nonzero for all neighboring sites, and the excitation energies $E_i^{(\ell)}$ of the $i$th Gutzwiller excited state at each site $\ell$. The remaining terms are the matrix elements of the non-local products of local operators

\begin{align}
{F}^{(\ell,\ell')}_{i_1,i_2,j_1,j_2} &= {B}^{(\ell)*}_{j_1,i_1} {B}^{(\ell')}_{i_2,j_2} + {B}^{(\ell)}_{i_1,j_1} {B}^{(\ell')*}_{j_2,i_2} \\ \textrm{with}  \quad {B}^{(\ell)}_{i,j} &= \prescript{}{\ell}{\left\langle i \right|} \hat{b}_{\ell} \left| j \right\rangle_{\ell} - \phi_{\ell} \delta_{i,j}, \nonumber
\end{align}
where $\phi_{\ell}$ are the previously defined self-consistent mean-field values associated with the local annihilation operator.

\subsection{Diagonalization of $\mathcal{H}^{(2)}_{\textrm{QP}}$}\label{sec:QP_diagonalization}

In order to preserve the bosonic structure of the operators, the diagonalization of~\eqref{eq:H2_in_quasimomentum} has to be performed on the symplectic space, namely by diagonalizing $\Sigma \mathcal{H}^{(2)}_{\textrm{QP}}$, where $\Sigma = \begin{pmatrix} \mathbb{1}_{(N-1)L^2} & 0 \\ 0 & - \mathbb{1}_{(N-1)L^2} \\ \end{pmatrix}$. This yields the representation of $\hat{\mathcal{H}}^{(2)}_{\textrm{QP}}$ in terms of the generalized Bogoliubov-type QP modes 

\begin{align}
\beta_{\gamma} &\equiv \mathbf{x}^{(\gamma)^{\dag}} \Sigma \begin{pmatrix} {\boldsymbol{\sigma}} \\ {\boldsymbol{\sigma}}^{\dag} \end{pmatrix} \equiv \mathbf{u}^{(\gamma)^{\dag}}{\boldsymbol{\sigma}} + \mathbf{v}^{(\gamma)^{\dag}} {\boldsymbol{\sigma}}^{\dag}, \label{eq:QP_modeop1} \\
\beta_{\gamma}^{\dag} &\equiv -\mathbf{y}^{(\gamma)^{\dag}} \Sigma \begin{pmatrix} {\boldsymbol{\sigma}} \\ {\boldsymbol{\sigma}}^{\dag} \end{pmatrix} \equiv \mathbf{v}^{(\gamma)^{T}}{\boldsymbol{\sigma}} + \mathbf{u}^{(\gamma)^{T}} {\boldsymbol{\sigma}}^{\dag}. \label{eq:QP_modeop2}
\end{align}
These are given by the eigenvectors of the eigenvalue equations $\Sigma {\mathcal{H}}_{\textrm{QP}} ^{(2)} \mathbf{x}^{(\gamma)} = \omega_{\gamma} \mathbf{x}^{(\gamma)}$ with $\mathbf{x}^{(\gamma)} = \left( \mathbf{u}^{(\gamma)} , -\mathbf{v}^{(\gamma)} \right)^{T}$ and $\Sigma {\mathcal{H}}_{\textrm{QP}} ^{(2)} \mathbf{y}^{(\gamma)} = -\omega_{\gamma}^* \mathbf{y}^{(\gamma)}$ with $\mathbf{y}^{(\gamma)} = \left( -\mathbf{v}^{(\gamma)} , \mathbf{u}^{(\gamma)} \right)^{\dagger}$. Thus all  
QP frequencies $\omega_{\gamma}$ appear in pairs and those with a nonzero imaginary part represent unstable QP modes \footnote{As an empirical observation, unstable QP modes are only encountered for MF states far from the ground state.}. By requiring the normalization condition $|\mathbf{u}^{(\gamma)}|^2 - |\mathbf{v}^{(\gamma)}|^2 = 1$ in analogy to regular Bogoliubov theory, we preserve the (approximate) bosonic commutation relations~\eqref{eq:GWfluc_commut_red} and~\eqref{eq:GWfluc_commut_zero}, so $\left[ \beta_{\gamma} , \beta^{\dag}_{\gamma'} \right] = \delta_{\gamma,\gamma'}$. We note that the two halves of the eigenvectors $\mathbf{v}^{(\gamma)}$ and $\mathbf{u}^{(\gamma)}$ can be interpreted as dual wave functions associated with particle and hole type fluctuations, respectively. As we will discuss in the following sections, these generalized Bogoliubov quasiparticles can be extended (with an associated lattice momentum, see Figs.~\ref{fig:QP_spec_W1} and~\ref{fig:QP_spec_W5}), localized \cite{Ros2015} or posses a finite core (see Fig.~\ref{fig:amp_decay_U20W10} and Sec.~\ref{sec:QP_loc_len}).

In the presence of a condensate one encounters a degenerate two-dimensional subspace constituted by an identity of the energy pair $\omega_{\gamma} = - \omega_{\gamma}^* = 0$, an expression which becomes numerically exact only for $N \rightarrow \infty$. In the case of an exact degeneracy the eigenvalue equation becomes $\Sigma \mathcal{H}^{(2)}_{\textrm{QP}} \mathbf{p} = 0$ and can be solved by an eigenvector of the form $\mathbf{p} = (\mathbf{u}^{(0)} , -\mathbf{u}^{(0)^*})^T$. In order to complete the representation of this two-dimensional subspace one has to introduce a second vector $\mathbf{q}$ within this subspace, which is best defined implicitly via $\Sigma \mathcal{H}^{(2)}_{\textrm{QP}} \mathbf{q} = -i \mathbf{p} / \tilde{m}$, where $\tilde{m}$ is a mass-like scalar. Therefore, we obtain two different operators taking the places of the Bogoliubov-like operators~\eqref{eq:QP_modeop1} and~\eqref{eq:QP_modeop2} for the doubly degenerate mode (these are discussed in further detail in \cite{Bissbort2012}):

\begin{align}
\mathcal{P} &\equiv \mathbf{p}^{\dag} \Sigma \begin{pmatrix} \tilde{\boldsymbol{\sigma}} \\ \tilde{\boldsymbol{\sigma}}^{\dag} \end{pmatrix} = \begin{pmatrix} \mathbf{u}^{(0)} \\ -\mathbf{u}^{(0)^*} \end{pmatrix}^{\dag} \Sigma \begin{pmatrix} \tilde{\boldsymbol{\sigma}} \\ \tilde{\boldsymbol{\sigma}}^{\dag} \end{pmatrix},  \label{eq:QP_modeop_P} \\
\mathcal{Q} &\equiv -\mathbf{q}^{\dag} \Sigma \begin{pmatrix} \tilde{\boldsymbol{\sigma}} \\ \tilde{\boldsymbol{\sigma}}^{\dag} \end{pmatrix} \equiv - i \begin{pmatrix} \mathbf{v}^{(0)} \\ \mathbf{v}^{(0)^*} \end{pmatrix}^{\dag} \Sigma \begin{pmatrix} \tilde{\boldsymbol{\sigma}} \\ \tilde{\boldsymbol{\sigma}}^{\dag} \end{pmatrix}.  \label{eq:QP_modeop_Q}
\end{align}
We note that $\mathcal{P}$ is a momentum-like operator that can be considered as the generator of translations in the global phase of the condensate mode \cite{Lewenstein1996}, so it represents the free motion of the complex phase factor of the condensate.

As a result of the (approximately) exact commutation relations of the QP mode operators, the second order Hamiltonian $\mathcal{H}^{(2)}_{\textrm{QP}}$ generally has the form

\begin{align}
\mathcal{H}^{(2)}_{\textrm{QP}} \approx \tilde{\sum_{\gamma}} \omega_{\gamma} \beta_{\gamma}^{\dag} \beta_{\gamma} + \frac{\mathcal{P}^2}{2\tilde{m}} + \frac{1}{2}\left( \tilde{\sum_{\gamma}} \omega_{\gamma} - \textrm{Tr}(h) \right). \label{eq:HQP_diag_full}
\end{align}
This representation is given in terms of the generalized Bogoliubov creation (annihilation) operators $\beta_{\gamma}^{\dag}$ ($ \beta_{\gamma}$) where the notation $\tilde{\sum}_{\gamma}$ represents the fact that the $\gamma = 0$ term in the sum is to be replaced by $\mathcal{P}$ whenever a condensate is present for $N \rightarrow \infty$. Otherwise, for small $N$, the lowest mode remains gapped such that the $\mathcal{P}$ term can be replaced by $\omega_0\beta^{\dag}_0\beta_0$. As all $\omega_{\gamma}>0$ the form~\eqref{eq:HQP_diag_full} implies that the quasiparticle ground state is characterized by $\langle \psi_{\textrm{QP}} | \beta^{\dag}_{\gamma} \beta_{\gamma} | \psi_{\textrm{QP}} \rangle = 0$ ($\langle \psi_{\textrm{QP}} | \mathcal{P}^2 | \psi_{\textrm{QP}} \rangle = 0$), so we can use $\beta_{\gamma} | \psi_{\textrm{QP}} \rangle = 0$ ($\mathcal{P} | \psi_{\textrm{QP}} \rangle = 0$) for all $\gamma$ as its implicit definition. This allows for the \textit{a posteriori} check of the central FOE approximation discussed in Sec.~\ref{sec:Commutator_deviation}. Regarding the spectral properties discussed in Sec.~\ref{sec:spec_func}, consideration of $\mathcal{P}$ and $\mathcal{Q}$ only yields a sub-leading [even self-canceling for the spectral function at $\omega=0$, see Sec.~\ref{sec:spec_func}] correction in the thermodynamic limit \cite{Frerot2016}, so we may neglect both for our purposes.
By expressing $\mathcal{H}^{(2)}_{\textrm{QP}}$ with $\beta_{\gamma}^{\dag}$ and $ \beta_{\gamma}$ in normal order we find a further scalar contribution proportional to $\tilde{\sum}_{\gamma} \omega_{\gamma}$. Note that both scalar terms generate a shift of the total energy. While both contributions $\tilde{\sum}_{\gamma} \omega_{\gamma}$ and $\textrm{Tr}(h)$ would diverge individually without truncation ($N \rightarrow \infty$), even in a finite system, in combination they only yield a finite correction of the quasiparticle ground-state energy. They effectively lower the energy of $| \psi_{\textrm{QP}} \rangle$ in relation to the energy of the \ac{MF} state $| \psi_{\textrm{MF}} \rangle$ as a result of an average down shift of the QP mode energies in relation to the energies of the Gutzwiller excitations.

This concludes the diagonalization of the disordered BHM up to second order in the Gutzwiller operators. The obtained generalized Bogoliubov modes can be of varying character. Either they behave like extended Bogoliubov quasiparticles with a well defined lattice momentum $\mathbf{k}$, as will be discussed in Sec.~\ref{sec:spec_func}, or they are localized at random sites with an exponential tail far from the center, as will be discussed in Sec.~\ref{sec:FOE-charact_qp_spec}. These two regimes correspond to low and high energy \ac{QP} excitations, respectively, which are separated by the \ac{ME} previously determined in \cite{Geissler2019a} and confirmed via the diverging localization length of the \ac{QP} excitations in Sec.~\ref{sec:FOE-charact_qp_spec}.

We note that the obtained representation is reminiscent of the emergent \ac{LIOM}s predicted within the \ac{MBL} phase \cite{Serbyn2013a,Huse2014,Ros2015,Chandran2015a,OBrien2016,Imbrie2017}, albeit on the lowest order of approximation where all coupling terms between the approximate \ac{LIOM}s are disregarded. Thus this parallel is expected to hold especially for strongly localized \ac{QP} states, where we consider the \ac{FOE} to yield a representation in terms of approximate \ac{LIOM}s for which the corresponding \ac{QP} states have very small localization lengths. This is indeed the case, as we show in the following section where we characterize the corresponding spectrum via its energies and the spatial localization of the \ac{QP} eigenstates. With regards to further studies these \ac{QP} mode operators may thus serve as an ideal starting point for the construction of proper \ac{LIOM}s, for example using methods discussed in \cite{Mierzejewski2018,Mierzejewski2020}.

\section{Characterization of the QP spectrum} \label{sec:FOE-charact_qp_spec}

In this section we extend our discussion beyond the ground state by considering and characterizing the general \ac{QP} fluctuations obtained within the \ac{FOE} method discussed in the previous Sec.~\ref{sec:FOE}. On the one hand, we analyze the distribution of the \ac{QP} energy levels and their gap statistics. On the other hand, we specifically discuss the exponential localization of the \ac{FOE} wave functions associated with the \ac{QP} excitations. Both aspects can be summarized in terms of two simple and fundamentally different measures related to localization. These are (i) the QP energy level spacing ratio $r \in [0,1]$ and (ii) the multi-fractality dimension $D \in [0,1]$ for the second moment of the \ac{QP} fluctuation wave functions. They reveal and allow for an independent characterization of the many-body \ac{ME} as discussed in detail in \cite{Geissler2019a}. But before doing so we first asses the validity of assuming negligible interactions between the Gutzwiller excitations for individual \ac{QP} modes.

\subsection{Gutzwiller population of \ac{QP} states}

\begin{figure}[t]
 \centering
 \includegraphics[width=0.99\columnwidth]{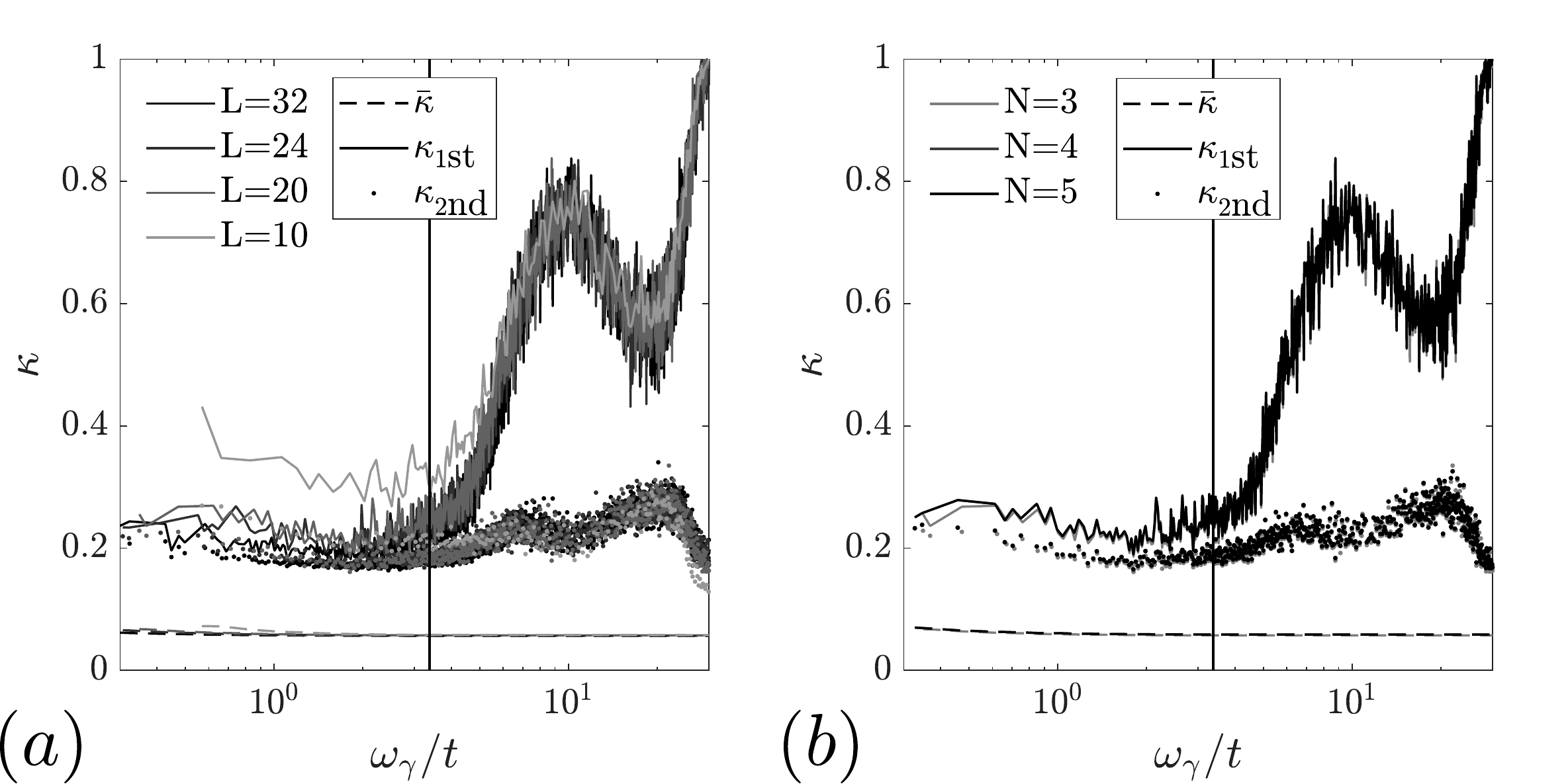}
 \caption{
 Disorder averaged fraction of local fluctuations $\kappa$ for the \ac{QP} excitations as a function of their disorder average energy $\omega_{\gamma}$. The vertical lines mark the inverse variance weighted mean of the \ac{ME} as determined in \cite{Geissler2019a} As given in the legend, the lattice average $\bar{\kappa}$ and the disorder average of the (2nd to) maximum ($\kappa_{2\textrm{nd}}$) $\kappa_{1\textrm{st}}$ are shown for fixed $U/t = 20$ and $W/t=5$. In $(a)$ $N=3$ and $L\in\lbrace 10,20,24,32 \rbrace$, while $L=20$ and $N\in\lbrace 3,4,5 \rbrace$ in $(b)$ as specified in the respective legends. For all cases $N_r=25$, except for $L=10$ where $N_r=50$.
 }
 \label{fig:kappa_QP_states}
\end{figure}

Similar to the \textit{a posteriori} check of $\kappa \ll 1$ for the \ac{QP} ground state in Sec.~\ref{sec:Commutator_deviation} we now perform the corresponding analysis for individual \ac{QP} excitations $\beta^{\dagger}_{\gamma} | \psi_{\textrm{QP}} \rangle$ of the \ac{QP} ground state. We focus on the representative parameters $U/t = 20$ and $W/t = 5$. For each realization the local Gutzwiller populations for a \ac{QP} excitation $\gamma$ and site $\ell$ are given by

\begin{align}
\langle \kappa_{\ell} \rangle_{\gamma} =& \langle \psi_{\textrm{QP}} | \beta_{\gamma} \sum_{i>0} \sigma^{(i)^{\dagger}}_{\ell} \sigma^{(i)}_{\ell} \beta_{\gamma}^{\dagger} | \psi_{\textrm{QP}} \rangle \\
=& \sum_{i>0} \left( \sum_{\alpha>0} |v_{i,\ell}^{(\alpha)}|^2 + |v_{i,\ell}^{(\gamma)}|^2 + |u_{i,\ell}^{(\gamma)}|^2 \right). \nonumber
\end{align}
To quantify the assumption and its limits we consider the average Gutzwiller population $\bar{\kappa}$ as well as the disorder average of the (2nd-to) maximum Gutzwiller population ($\kappa_{2\textrm{nd}}$) $\kappa_{1\textrm{st}}$. Given the site $\ell^{(\gamma)}_m$ of the maximum Gutzwiller population for each QP state and realization with $\langle \kappa_{\ell^{(\gamma)}_m} \rangle_{\gamma} > \langle \kappa_{\ell} \rangle_{\gamma}$ for every site $\ell$, these are defined as

\begin{align}
\bar{\kappa}(\omega_{\gamma}) =& \left\langle \sum_{\ell} \frac{\langle \kappa_{\ell}\rangle_{\gamma}}{L^2} \right\rangle_d, \\
\kappa_{1\textrm{st}}(\omega_{\gamma}) =& \left\langle \langle \kappa_{\ell^{(\gamma)}_m}\rangle_{\gamma} \right\rangle_d, \\
\kappa_{2\textrm{nd}}(\omega_{\gamma}) =& \left\langle \max_{\ell \in L^2 \setminus \ell^{(\gamma)}_m} \langle \kappa_{\ell}\rangle_{\gamma} \right\rangle_d.
\end{align}
Here, the lowest mode $\gamma=0$ is discarded as it is sub-leading in the thermodynamic limit (see Sec.~\ref{sec:QP_diagonalization}). 

These disorder averages with at least $N_r=25$ realizations are shown in Fig.~\ref{fig:kappa_QP_states}, either for fixed $N=3$ and $L \in \lbrace 10,20,24,32 \rbrace$ or for fixed $L=20$ and $N\in \lbrace 3,4,5 \rbrace$ using identical disorder realizations for each $N$. The former shows that there are strong finite size effects for very small systems, especially for $L=10$, while the latter shows that even a low truncation of $N=3$ is sufficient for a good convergence. Compared to the QP ground state the average fraction of local excitations $\tilde{\kappa} \approx 0.06$ is only slightly increased in any QP mode. The maximum population $\kappa_{1\textrm{st}}$ on the other hand is well below 1 for low energy states, but $\kappa_{1\textrm{st}}$ increases considerably for modes above the \ac{ME} $\omega_{\gamma} /t \gtrsim 3.4$ approaching 1 for energies $\omega_{\gamma} > U$. But these modes are very localized as the 2nd-to maximum $\kappa_{2\textrm{nd}}$ is nearly constant on either side of the \ac{ME}. Thus, the QP modes above the \ac{ME} are highly localized fluctuations. Even in the presence of QP modes the condition $\kappa < 1$ is thus typically fulfilled for excitation energies $\omega_{\gamma} < U$ while sizable interactions between QP modes ($\beta^{\dagger}_{\gamma}$) become most relevant for either a large number of extended fluctuations or nearby pairs of localized fluctuations (see also Sec.~\ref{sec:QP_loc_len}).

\begin{figure*}[t]
 \centering
 \includegraphics[width=0.9\textwidth]{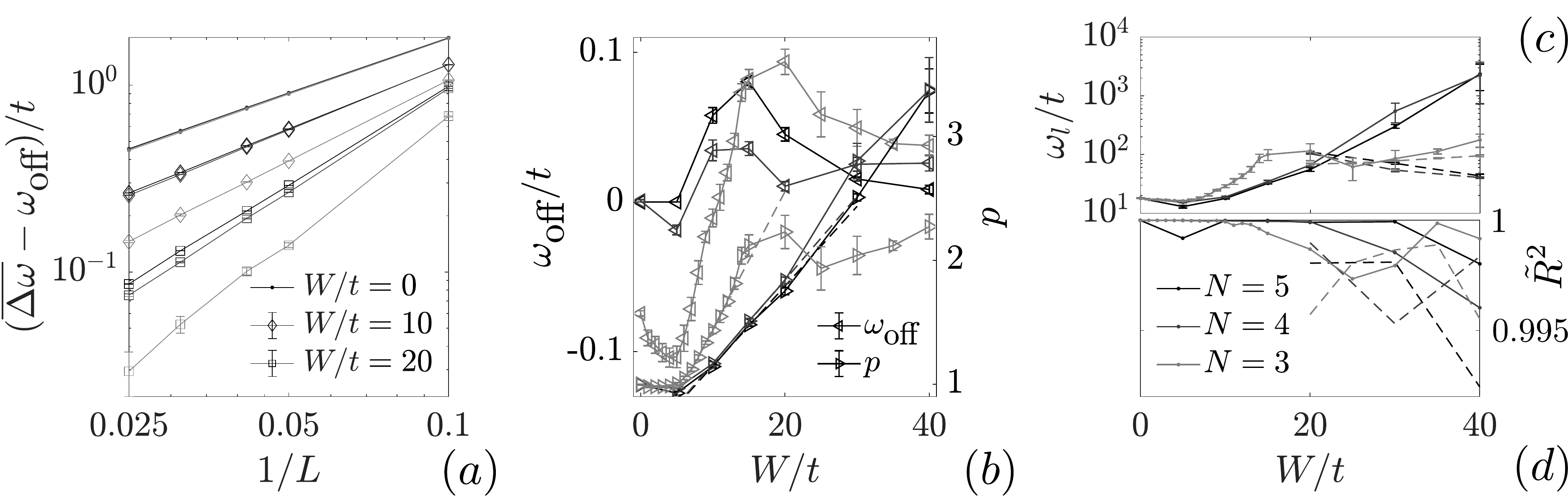}
 \caption{
Scaling of the lowest \ac{QP} gaps $\Delta \omega$ with the grayscale indicating the truncation $N = 3,4,5$, respectively, as given in the legend of $(d)$. $(a)$ Exemplary data for different values of $W/t$ given in the legend. Best fit parameters of the scaling ansatz eq.~\eqref{eq:QPgap_scaling} including one standard deviation in the error of the fit are given in $(b)$ (see legend) and $(c)$. Dashed lines in $(b)$ are guides to the eye. $(d)$ shows the adjusted parameter of convergence $\tilde{R}^2$ of the fits. Dashed lines in $(c)$ and $(d)$ correspond to fitting results for fixed $p=2$.
 }
 \label{fig:QP_gap_scaling}
\end{figure*}

\subsection{Superfluid vs. Bose-glass gap scaling}

Before we discuss the \ac{QP} spectrum we first take into account the lowest \ac{QP} excitations only, in order to discuss their relation to the ground state. To do so we consider the $n$ lowest \ac{QP} excitations $\omega_{\gamma}$ with $\gamma \leq n$. We note that for the local basis truncation $N \rightarrow \infty$ (see Sec.~\ref{sec:FOE_Gops}) in the presence of a \ac{MF} condensate follows $\omega_0 \rightarrow 0$ in which case this mode actually has to be represented by the momentum-like operator $\mathcal{P}$, as discussed in Sec.~\ref{sec:FOE_qpHamil}. Thus the lowest relevant average $n$-gaps are given by $\Delta \omega_{\gamma} = \langle \omega_{\gamma} - \omega_0 \rangle_d$. In a superfluid it is well known that the lowest energy excitations are Goldstone modes following a linear dispersion relation. Irrespective of the spatial dimension the smallest possible lattice momenta on an isotropic lattice have $|\mathbf{k}_{\textrm{min}}| = \pi/La, \sqrt{2}\pi/La,\ldots$ implying $\Delta \omega_{\gamma} \propto 1/L$ for sufficiently small $\gamma$. In contrast, for strong disorder excitations are expected to be increasingly uncorrelated such that the average level spacing becomes inversely proportional to the total number of levels. Therefore, we expect $\Delta \omega_{\gamma} \propto 1/L^2$ for sufficiently strong disorder as the number of \ac{QP} modes within the \ac{FOE} is proportional to the number $L^2$ of lattice sites. As we are only interested in the scaling with $L$ it is numerically beneficial to consider the average of the 8 lowest gaps $\overline{\Delta \omega} \equiv \sum_{\gamma=1}^8\Delta \omega_{\gamma} / 8 $ corresponding to the longest wavelength modes $|\mathbf{k}| \in \lbrace \pi/La, \sqrt{2}\pi/La \rbrace$ of the superfluid. For this average we assume the following generic scaling.

\begin{align}
\overline{\Delta \omega} = \frac{\omega_l}{L^p} + \omega_{\textrm{off}} \label{eq:QPgap_scaling}
\end{align}
Here, the first term represents the system size scaling with some power $p$ and  an effective local single-site gap $\omega_l$ while $\omega_{\textrm{off}}$ is an offset energy. These parameters are determined via fitting. We perform this scaling for $U/t = 20$, $W/t \in \left[ 0 , 40 \right]$ and $N\in \lbrace 3,4,5 \rbrace$ [corresponding to light grey, dark grey and black in Fig.~\ref{fig:QP_gap_scaling}]. Exemplary data for $W/t \in \lbrace 0, 10, 20 \rbrace$ is shown in Fig.~\ref{fig:QP_gap_scaling}$(a)$ with errors of the mean from the disorder sampling. The obtained values for $\omega_{\textrm{off}}$, $p$ and $\omega_l$ are given in Figs.~\ref{fig:QP_gap_scaling}$(b,c)$, while the corresponding adjusted coefficient of determination $\tilde{R}^2$ is shown in Fig.~\ref{fig:QP_gap_scaling}$(d)$. All fits are nearly exact with an adjusted parameter of convergence $\tilde{R}^2 \approx 1$. 

Regarding $p$, a truncation $N>3$ is sufficient to determine this scaling exponent in the vicinity of the \ac{SF} to \ac{BG} transition for the considered system sizes (see Fig.~\ref{fig:QP_gap_scaling}), although one has to be careful for disorder $W/t>20$ \cite{Geissler2019a}. Just as expected we find $p=1$ for sufficiently weak disorder consistent with a \ac{SF} phase while the exponent increases approximately linearly beyond 1 above a critical disorder $W_{c,\Delta\omega}$. Linear fits (dashed lines) in Fig.~\ref{fig:QP_gap_scaling}$(b)$ cross $p=1$ at $W_{c,\Delta \omega} = 5.98(8),7.5(2),7.7(7)$ corresponding to $N=3,4,5$, respectively, and thus well above the \ac{MF} result. Regarding the effective single site gap $\omega_l \approx U$ for sufficiently weak disorder, as one would expect in the single site limit. Notably, in the opposite limit at strong disorder $W > 20t$ ($W > U$) we find best fits with $p>2$ and $\omega_l \gg U,W$ which also have the lowest fit quality [see Figs.~\ref{fig:QP_gap_scaling}$(b)$, $(c)$ and $(d)$]. As such a runaway effective local gap seems unphysical, we also assume a fixed value $p=2$ as discussed earlier for $W \geq 20 t$. We then find fits of nearly identical quality [dashed lines in Fig.~\ref{fig:QP_gap_scaling}$(d)$] but with much lower effective local gaps $\omega_l$ [dashed lines in Fig.~\ref{fig:QP_gap_scaling}$(c)$]. 

In conclusion, we find a finite-size scaling of lowest \ac{QP} excitations consistent with a dissolving spectrum of Goldstone modes for increasing disorder, as expected for a \ac{SF} to \ac{BG} transition. Furthermore, the behavior of the scaling exponent $p$ remains unclear at strong disorder $W > 20t$ ($W > U$) where in earlier works we have shown the need for even greater truncation $N>5$ to obtain converged lowest energy \ac{QP} excitations \cite{Geissler2019a}.

\subsection{Level spacing statistics} 
\begin{figure}[t]
 \centering
 \includegraphics[width=0.99\columnwidth]{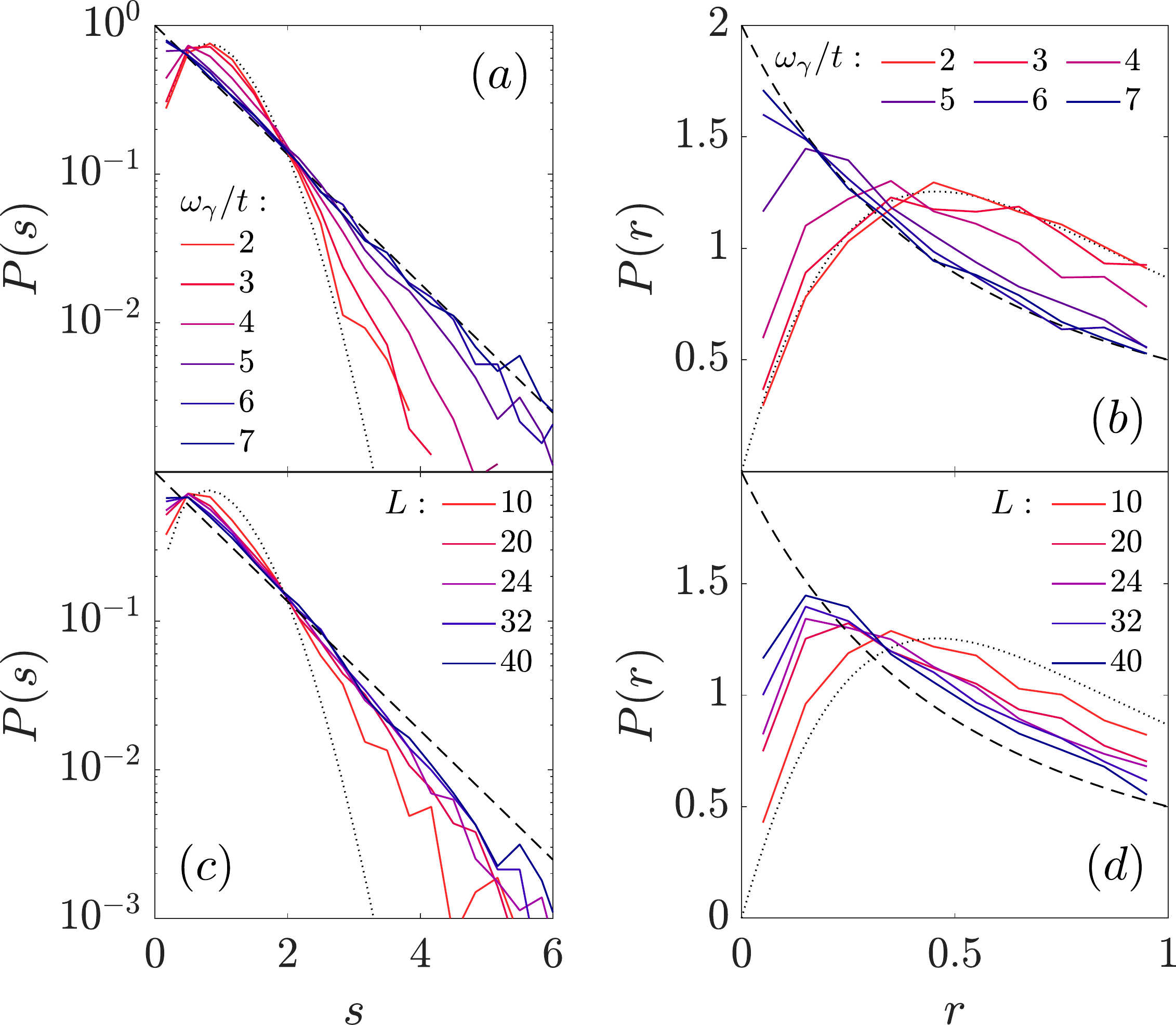}
 \caption{
 Gap and Gap ratio distributions $P(s)$ $(a,c)$ and $P(r)$ $(b,d)$, respectively, for the QP spectrum of~\eqref{eq:MBL-hamil} for $U/t=20$ and $W/t=5$. For reference the analytic predictions related to Poisson statistics (dashed lines) and for the \ac{goe} (dotted lines) are included in all plots. In $(a,b)$ each distribution covers an energy window $\omega_{\gamma}$ of width $t$ centered at various $\omega_{\gamma}/t \in [2,7]$ (see legend), with $L = 40$. For $(c,d)$ the central energy is fixed to $\omega_{\gamma}/t = 5$, while the linear system size $L$ is varied (see legend).
 }
 \label{fig:Gap_distribs}
\end{figure}

Next, we focus on the \ac{QP} spectrum beyond the low energy regime. To characterize an \ac{MBL}-like transition the gap ratio $r=r_{\gamma}$ is the most prevalent measure, which in terms of the \ac{QP} energy gaps $\Delta \omega_{\gamma} = \omega_{\gamma+1} - \omega_{\gamma}$ we define as 
\begin{align}
r_{\gamma} \equiv \left\langle \frac{ \textrm{min}[\Delta \omega_{\gamma-1},\Delta \omega_{\gamma}] }{ \textrm{max}[\Delta \omega_{\gamma-1},\Delta \omega_{\gamma}]} \right\rangle_d.
\end{align}
The statistical properties of $r$ and the rescaled level spacing $s \equiv \Delta \omega_{\gamma} / \overline{\Delta \omega_{\gamma}}$, where $\overline{\Delta \omega_{\gamma}}$ is the mean level spacing, are well known from random matrix theory \cite{Atas2013,Sierant2018}. In the delocalized regime the respective probability distributions $P(s)$ and $P(r)$ are determined by the \ac{goe}, while in the localized regime these follow from \ac{P} statistics. The former case is well described by Wigner's surmise $P_W(s) = \frac{\pi}{2} s \exp (-\frac{\pi}{4}s^2)$ with 
\begin{align}
P_W(r) = \frac{27}{4} \frac{r+r^2}{(1+r+r^2)^{5/2}},
\end{align}
while the Poissonian has the simple form $P(s) = \exp(-s)$ with $P(r) = 1/(1+r)^2$. In Fig.~\ref{fig:Gap_distribs} we show distributions obtained for the QP spectra at $U/t = 20$ and $W/t = 5$ in the vicinity of the low energy \ac{ME} at about $\omega_{\gamma}/t \approx 3.5$. While the spectra at low $\omega_{\gamma}$ reproduce the \ac{goe} prediction, the distributions approach \ac{P} behavior for increased energies [see Fig.~\ref{fig:Gap_distribs}$(a,b)$], consistent with crossing a \ac{ME} somewhere in between. If, on the other hand, we increase the system size while keeping the energy window fixed to $\omega_{\gamma}/t=5\pm0.5$ [see Fig.~\ref{fig:Gap_distribs}$(c,d)$] we again find that the distributions interpolate from near \ac{goe} to \ac{P}-like behavior. This finite-size scaling behavior is consistent with QP states that are on the localized side of the \ac{ME}.

\begin{figure}[t]
 \centering
 \includegraphics[width=0.70\columnwidth]{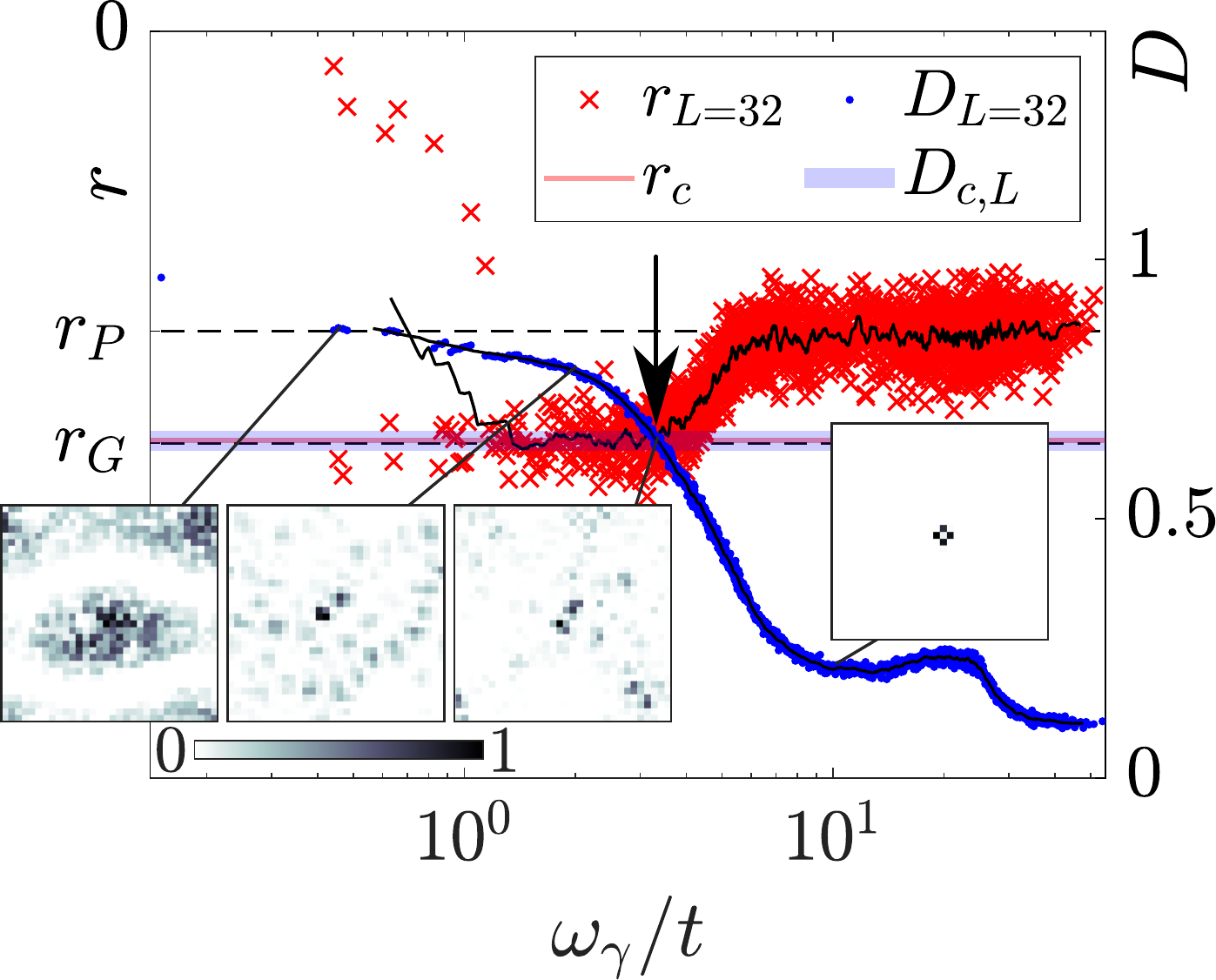}
 \caption{
 Gap ratio $r$ (left ordinate) and fractal dimension $D$ data (right ordinate) as functions of the QP energy $\omega_{\gamma}/t$ for $U/t = 20$, $W/t=5$ and $L=32$ averaged over 95 realizations. Black lines are moving averages as a guide to the eye and dashed lines mark $r_P$ and $r_{G}$. The crossing point (vertical arrow) of the data with the critical $r_c$ (shaded red, narrow) and $D_c$ (shaded blue, wide) mark the ME. \textit{Insets}: Exemplary squared QP wave functions $|\mathbf{v}_{\ell}^{(\gamma)}|^2$ normalized to the maximum. Reproduced from \cite{Geissler2019a}.
 }
 \label{fig:r_D_sample_U20W5}
\end{figure}

Random matrix theory furthermore predicts the expectation value of $r$ within each ensemble to $ r_{\textrm{G}} \approx 0.5307$ and $ r_{\textrm{P}}= 2 \textrm{ln} 2 - 1 \approx 0.3863$ for the \ac{goe} and \ac{P} statistics, respectively \cite{Atas2013}. In Fig.~\ref{fig:r_D_sample_U20W5} we show $r$ as a function of the \ac{QP} energies $\omega_{\gamma}$. For sufficiently low energies most $r_{\gamma} \approx r_{\textrm{G}}$, as expected for non-localized states. Outliers towards extremely small values result from a systematic finite-size effect. For not too strong disorder, such as $W/U=0.25$ in this case, the low energy part of the QP spectrum is only weakly disturbed, as visible by the nearly plane wave character of the wave function in the first inset of Fig.~\ref{fig:r_D_sample_U20W5}. Thus one finds clusters of near-degenerate \ac{QP} excitations in the spectrum for which the lattice momentum $\mathbf{k}$ still is a good approximate quantum number. The number of states in each cluster is related to the underlying $90^{\circ}$ rotational and reflection symmetries of the corresponding disorder-free \ac{QP} excitation bands, as can be seen in a clustering of the fractal dimension $D=D^{(\gamma)}_L$ of the QP fluctuation wave functions $\mathbf{v}^{(\gamma)}$ (see Fig.~\ref{fig:r_D_sample_U20W5}, blue dots) which we discuss in the following. 

\subsection{Decay of fluctuation wave functions}\label{sec:QP_loc_len}
\begin{figure}[t]
 \centering
 \includegraphics[width=0.95\columnwidth]{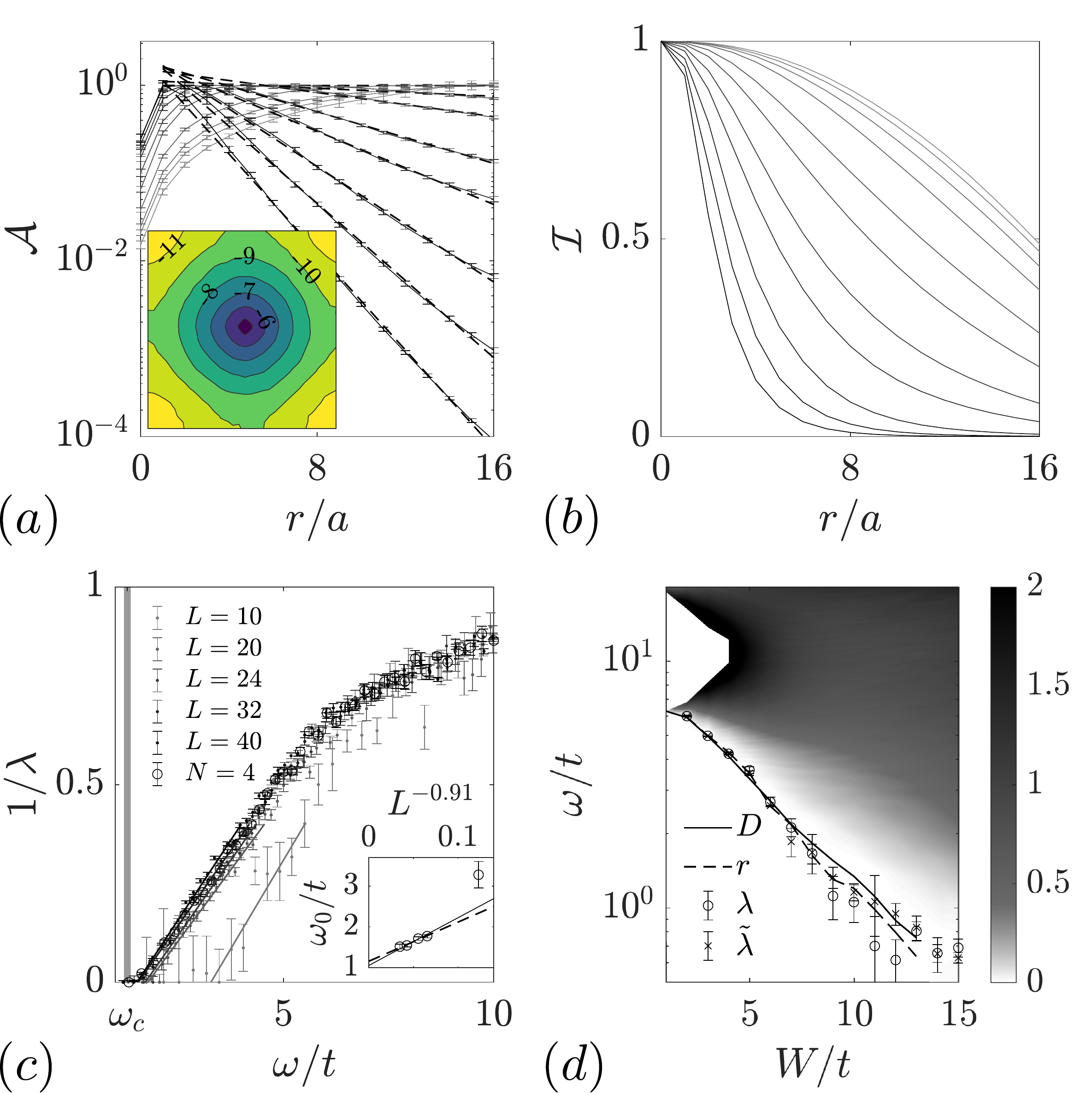}
 \caption{
 Decay behavior of $|\mathbf{v}^{(\gamma)}_{\ell}|^2$ for $U/t = 20$ in~\eqref{eq:MBL-hamil} with $L=32$ and $N=3$ if not specified otherwise. For $W/t = 7$ examples of $\mathcal{A}(r)$ and $\mathcal{I}(r)$ are shown in $(a)$ and $(b)$. All distributions are averaged over bins of 16 levels closest in energy to $\omega/t \in \left[ 0.75 , 5.75 \right]$ increasing in steps of $0.5$ from light gray to black (top to bottom at large distances). Inset in $(a)$ shows a corresponding typical state $ \left\langle \log_{10} |\mathbf{v}^{(\gamma)}_{\ell}|^2 \right\rangle_d$ at $\omega/t = 5.5$. Dashed lines in $(a)$ are exponential fits of eq.~\eqref{eq:decay_ansatz} and corresponding inverse decay lengths $1/\lambda$ are given in $(c)$ for $W/t = 10$ and $W/t \in \left[ 1, 15 \right]$ in $(d)$. Data sets in $(c)$ for $L \in [10,20,24,32,40]$ are accompanied by linear fits (solid lines) and circles have $L=32$ and $N=4$ (see legend). Zeros of these fits are given in the inset with solid (dashed) linear fit lines for all $L$ ($L>10$). $\omega_c$ in $(c)$ and lines in $(d)$ represent the \ac{ME} determined in \cite{Geissler2019a} via fractal dimension $D$ or gap ratio $r$ (see legend), while circles (crosses) are derived from the decay length $\lambda$ ($\tilde{\lambda}$) for all sizes ($L>10$).   
 }
 \label{fig:amp_decay_U20W10}
\end{figure}

While the discussed level statistics are fully consistent with a \ac{ME} in the disordered \ac{BHM} we now consider the localization properties of the fluctuation wave function $|\mathbf{v}^{(\gamma)}_{\ell}|^2 = \sum_{i>0} |\mathbf{v}^{(\gamma)}_{\ell,i}|^2$ directly. Here, we analyze the typical radial wave function amplitude $\mathcal{A}(r)$ oriented at its center-of-mass $\mathbf{r}_0$ for each level and disorder realization. The most relevant notion of distance is given by the minimal number of links between two sites. Thus, we define the norm $|\cdot|$ of a lattice vector $\mathbf{r}$ via its spatial components $x$ and $y$ as
\begin{align}
 \left| \mathbf{r} \right| = \sum_{i=x,y} |\mathbf{r}_i|
\end{align}
while we consider the center-of-mass $\mathbf{r}_0 = \lfloor \sum_{\ell} \mathbf{r}_{\ell} |\mathbf{v}^{(\gamma)}_{\ell}|^2 / \sum_{\ell} |\mathbf{v}^{(\gamma)}_{\ell}|^2 \rfloor$ with $\lfloor \cdot \rfloor$ denoting a rounding to the nearest site. Using these we define
\begin{align}
\tilde{\mathcal{A}}_{\gamma}(r) \equiv & \sum_{\lbrace \ell \mid \left| \mathbf{r}_{\ell} - \mathbf{r}_0 \right| = r \rbrace} \exp \left[ \left\langle \log |\mathbf{v}^{(\gamma)}_{\ell}|^2 \right\rangle_d \right],\\
\tilde{\mathcal{I}}_{\gamma}(r) \equiv & \sum_{r'>r} \tilde{\mathcal{A}} (r')
\end{align}
giving the angular integral of the typical wave function $\tilde{\mathcal{A}}_{\gamma}(r)$ and its radial integral $\tilde{\mathcal{I}}_{\gamma}(r)$. For convenience we scale either by its respective maximum: $\mathcal{A}_{\gamma}(r) = \tilde{\mathcal{A}}_{\gamma}(r) / \textrm{max}_{r}\left( \tilde{\mathcal{A}}_{\gamma}(r) \right)$ and the latter by the full sum: $\mathcal{I}_{\gamma}(r) = \tilde{\mathcal{I}}_{\gamma}(r)/\tilde{\mathcal{I}}_{\gamma}(0)$. For $U/t = 20$, $W/t = 7$ and $L=32$ a few examples of both are shown in Fig.~\ref{fig:amp_decay_U20W10}$(a)$ and $(b)$, respectively. Above a certain energy all \ac{QP} wave functions decay exponentially which also implies an exponential suppression of the third and fourth order interactions between these localized \ac{QP} excitations as a function of the distance between the respective center-of-mass of the involved \ac{QP} modes. Such a behavior is also expected for the \ac{LIOM}s commonly considered to describe \ac{MBL}.

In the following, energies are binned over consecutive energy levels so we discard the level index from here on and instead consider the mean energies of the bins. The behavior of either one is consistent with our findings so far. Note the deviations from a circular shape of the contour lines for a typical state $\langle\log_{10}|\mathbf{v}_{\ell}|^2 \rangle_d$, depicted in the inset of Fig.~\ref{fig:amp_decay_U20W10}$(a)$ for $\omega/t = 5.5$, justifying our distance definition. At this moderate disorder the fluctuation wave functions become strongly localized above some \ac{QP} energy $\omega_c$ associated with the mobility edge as visible by the exponential decay of $\mathcal{A}$ and $\mathcal{I}$ for large disorder. Especially the behavior of $\mathcal{I}$ at high energies implies that the majority of \ac{QP} state is constrained to the sites close to some central site, while the examples given in Fig.~\ref{fig:r_D_sample_U20W5} show that of these sites usually only a few actually contribute. To quantify the decay of the \ac{QP} wave-function we consider the following ansatz for the tail of $\mathcal{A}(r)$ fitted up to $r \leq L/2$:

\begin{align}
\mathcal{A}(r) \approx \exp \left( - \frac{r}{\lambda} + \xi \right). \label{eq:decay_ansatz}
\end{align}
Its parameters are an irrelevant offset $\xi$ related to the onset of the tail and the decay length $\lambda$. We always find $1/\lambda > 0$ (with an adjusted parameter of convergence that mostly is $\tilde{R}^2 > 0.99$) for any \ac{QP} excitation of sufficiently high energy -- that is, above the \ac{ME} -- in the thus localized part of the spectrum. We find this behavior for any local interaction $U$ and disorder $W$ which is also strongly convergent for sufficiently large $L>10$ and $N\geq3$. Thus, we can consider the inverse of the decay length as an order parameter, as $\lambda$ diverges at the transition from localized to extended states. Indeed, starting at high energies [see Fig.~\ref{fig:amp_decay_U20W10}$(c)$] or strong disorder [for sufficiently low energy, see Fig.~\ref{fig:amp_decay_U20W10}$(d)$] and lowering either the energy or the disorder strength, $1/\lambda$ eventually tends to zero within the resolvable states (limited by $N$ and $L$). This is nicely captured by linear fits for small $1/\lambda$ to 

\begin{align}
\frac{1}{\lambda} = a(\omega - \omega_0).
\end{align}
Here, $a$ is the slope with proper units and $\omega_0$ is the zero. Both terms are determined by fits for different $L$ [see Fig.~\ref{fig:amp_decay_U20W10}$(c)$ and inset]. The scaling of $\omega_0$ in turn follows a simple relation of the form

\begin{align}\label{eq:lambda_finit-size}
\omega_0 = \omega_c + \frac{\bar{\omega}}{L^{1/\nu}},
\end{align}
with the critical energy $\omega_c$ corresponding to the \ac{ME}, the rescaled energy $\bar{\omega}$ and the finite-size scaling exponent $1/\nu = 0.91(4)$ determined in our previous works \cite{Geissler2019a}. We note however that determining $\lambda$ for $\mathcal{A}(r)$ is problematic at small system sizes $L < 10$ and for the least localized \ac{QP} excitations which have a substantial inner region such that the onset of the decay is shifted outwards [see Fig.~\ref{fig:amp_decay_U20W10}$(a)$]. Then $\lambda$ may be overestimated for small $L$ and the least localized low energy excitations. This is visible in Fig.~\ref{fig:amp_decay_U20W10}$(c)$ where the deviation in $1/\lambda$ for $L=10$ and $L>10$ increases with $1/\lambda \rightarrow 0$. When fitting~\eqref{eq:lambda_finit-size} to determine $\omega_c$ we therefore distinguish two cases, one with $L=10$ included [solid line in the inset of Fig.~\ref{fig:amp_decay_U20W10}$(c)$ and labeled $\lambda$ in $(d)$] and the other with $L=10$ excluded [dashed line in the inset of Fig.~\ref{fig:amp_decay_U20W10}$(c)$ and labeled $\tilde{\lambda}$ in $(d)$].

Considering the drop in fit quality of~\eqref{eq:decay_ansatz} for $L=10$ due to the shift of the onset of decay and the weak decay when approaching the \ac{ME} from the localized side we find a \ac{ME} that closely overlaps with our earlier predictions, which relied on the finite-size scaling of the gap ratio and the fractal dimension of the \ac{QP} excitation states \cite{Geissler2019a}. For the sake of completeness we therefore finish this section with a brief discussion of the fractal dimension of the \ac{QP} wave-functions to show how the various observables related to the \ac{ME} compare.

\subsection{Fractal dimension of fluctuations} 
Analogous to the scaling of $q$-moments $R_q = \sum_{n} |\psi_{\nu}|^{2q}$ of many-body eigenstates where $n$ labels the partial amplitudes of a given many-body basis \cite{Hentschel1983,Mace2018,Lindinger2019}, our analysis is based on the local amplitudes of the wave function $|\mathbf{v}^{(\gamma)}_{\ell}|^2 = \sum_{i>0} |\mathbf{v}^{(\gamma)}_{\ell,i}|^2$ (and $q=2$):

\begin{align}
D \equiv D_L^{(\gamma)} = - \log_{L^2} \left[ \frac{\sum_{\ell}^{L^2} |\mathbf{v}^{(\gamma)}_{\ell}|^4}{\sum_{\ell}^{L^2} |\mathbf{v}^{(\gamma)}_{\ell}|^2} \right]. \label{eq:fractal_dim}
\end{align}
In contrast to many-body eigenstates the fluctuation wave function preserves real-space information in its amplitudes, so $D^{(\gamma)}_L$ characterizes the spatial extension of \ac{QP} fluctuations in relation to the system size (see insets in Fig.~\ref{fig:r_D_sample_U20W5}). As shown in our previous work $r$ and $D$ can be used to determine the \ac{ME} of the \ac{QP} spectrum [see Fig.~\ref{fig:amp_decay_U20W10}$(c,d)$] via the critical values $r_c(W)$ and $D_c(W)$ \cite{Geissler2019a}. As we have already seen for the level statistics, where high energy \ac{QP} excitations have \ac{P} statistics corresponding to localized states, also $D$ quickly tends to 0 above the \ac{ME} where the \ac{QP} states are centered at arbitrary sites and only involve a few of the nearest sites [see Figs.~\ref{fig:r_D_sample_U20W5} and~\ref{fig:amp_decay_U20W10}$(a,b)$]. We note that this behavior is very typical of \ac{LIOM}s implying that the \ac{QP} modes can be considered their lowest order approximation via the definition $I_{\gamma}^{(0)} = \beta^{\dagger}_{\gamma} \beta_{\gamma}$. In case of the existence of actual \ac{LIOM}s, corrections to this lowest order can be determined by the thus far neglected Hamiltonian terms $\mathcal{H}^{(3)}$ and $\mathcal{H}^{(4)}$, analogous to a weak coupling expansion \cite{Ros2015}. But as the \ac{FOE} is effectively a strong coupling expansion, already the lowest order goes beyond a single particle description. In particular, using the \ac{FOE} method the \ac{QP} ground state as well as its \ac{QP} excitations can be highly entangled as has been shown in the previous work \cite{Geissler2019a}.

In summary, we have shown that the localization of the \ac{QP} states is well characterized by the level spacing statistics, the decay length of the fluctuation wave-functions and their related fractal dimension. From these we obtain matching predictions of the \ac{ME}. Notably, a very similar inverted many-body \ac{ME} and \ac{MBL} transition has previously been found via exact methods for small one dimensional systems \cite{Sierant2018,Yao2020}. Furthermore, we have shown that the finite-size scaling of the lowest gaps is consistent with a \ac{SF} to \ac{BG} transition of the ground state. In the next section we will complete this picture by discussing various spectral functions of the quasiparticle ground state $| \psi_{\textrm{QP}} \rangle$.


\section{Spectral functions}\label{sec:spec_func}

As shown in Sec.~\ref{sec:QP_diagonalization}, one can derive a simple implicit definition for a corrected QP ground state $|\psi_{\textrm{QP}}\rangle$ by requiring the condition $\beta_{\gamma} |\psi_{\textrm{QP}} \rangle = 0$ for all QP modes $\gamma$ (and $\mathcal{P}|\psi_{\textrm{QP}} \rangle = 0$). Using this definition it is straightforward to determine the single particle spectral functions. Here, we focus on the normalized dynamic structure factor $\bar{S}(\mathbf{k},\omega)$ and the spectral function $\mathcal{A}(\mathbf{k},\omega) = -\textrm{sgn}(\omega)\textrm{Im}\left[ \sum_{\ell,\ell'} e^{-i\mathbf{k}\cdot(\mathbf{r}_{\ell} - \mathbf{r}_{\ell'})} G_{\ell \ell'}(\omega) \right]/L^2 \pi$ defined via the single-particle lattice Green's function $G_{\ell \ell'}(\omega)$. Using the notation $\langle \cdot \rangle_{\textrm{QP}} \equiv \langle \psi_{\textrm{QP}} | \cdot | \psi_{\textrm{QP}} \rangle$, their spectral representation for the \ac{QP} ground state can respectively be written as \cite{Geissler2018}

\begin{align}
\mathcal{A}(\mathbf{k},\omega) =& \theta(\omega) \mathcal{A}_{>}^{(2)}(\mathbf{k},\omega) - \theta(-\omega) \mathcal{A}_{<}^{(2)}(\mathbf{k},\omega) \label{eq:spec_func} \\
=& \theta(\omega) \langle \langle \hat{b}_{\mathbf{k}} \delta (\hat{H}^{(2)} - E_0 - \omega) \hat{b}_{\mathbf{k}}^{\dagger} \rangle_{\textrm{QP}} \rangle_d \nonumber \\
& - \theta(-\omega) \langle \langle \hat{b}_{\mathbf{k}}^{\dagger} \delta (\hat{H}^{(2)} - E_0 + \omega) \hat{b}_{\mathbf{k}} \rangle_{\textrm{QP}}\rangle_d, \nonumber \\ 
\bar{S}(\mathbf{k},\omega) =& \langle \langle \hat{n}_{\mathbf{k}} \delta (\hat{H}^{(2)} - E_0 - \omega) \hat{n}_{\mathbf{k}} \rangle_{\textrm{QP}}\rangle_d / N_p \label{eq:DSF}
\end{align}
with $\theta(\cdot)$ the Heaviside-Theta function, $E_0 = \langle \psi_{\textrm{QP}} | \hat{H}^{(2)} | \psi_{\textrm{QP}} \rangle $ the \ac{QP} ground state energy, $N_p$ the total number of particles and the label $d$ signifying the disorder average. The spectral function is defined in terms of the greater and lesser spectral functions $\mathcal{A}_{>}^{(2)}(\mathbf{k},\omega)$ and $\mathcal{A}_{<}^{(2)}(\mathbf{k},\omega)$ characterizing particle and hole excitations, respectively. 
Additionally, we consider the static counter parts, the momentum distribution $n(\mathbf{k}) = -\int_{- \infty}^{0} \mathcal{A}^{(2)}_{<}(\mathbf{k},\omega) d \omega$ and the static structure factor $S(\mathbf{k}) = \int_0^{\infty} \bar{S}(\mathbf{k},\omega)$. These are given in terms of Fourier transforms of the local creation, annihilation and number operators, 

\begin{align}
\hat{b}_{\mathbf{k}} = \frac{1}{\sqrt{L^2}} \sum_{\ell} e^{-i \mathbf{k}\cdot \mathbf{r}_{\ell}} \hat{b}_{\ell}, \\
\hat{b}_{\mathbf{k}}^{\dagger} = \frac{1}{\sqrt{L^2}} \sum_{\ell} e^{i \mathbf{k}\cdot \mathbf{r}_{\ell}} \hat{b}_{\ell}^{\dagger},\\
\hat{n}_{\mathbf{k}} = \sum_{\ell} e^{i \mathbf{k}\cdot \mathbf{r}_{\ell}} \hat{n}_{\ell}.
\end{align}
Furthermore, due to the completeness of each eigenbasis $\lbrace | i \rangle_{\ell} \rbrace$ of the local \ac{MF} Hamiltonians~\eqref{eq:MF-Hamil} any local operator $\hat{O}^{(\ell)}$ has an exact representation within this basis, in terms of the local Gutzwiller operators:

\begin{align}
\hat{O}^{(\ell)} =& \sum_{i,j \geq 0} {}_{\ell}\langle i | \hat{O}^{(\ell)} | j \rangle_{\ell} | i \rangle_{\ell} {}_{\ell}\langle j | \label{eq:local_OP_representation} \\
 \equiv & \sum_{i,j \geq 0}  O^{(\ell)}_{ij} | i \rangle_{\ell} {}_{\ell}\langle j | \nonumber \\
 =& \sum_{i > 0} \left( O^{(\ell)}_{i0} \sigma_{\ell}^{(i)^{\dagger}} + O^{(\ell)}_{0i} \sigma_{\ell}^{(i)} \right) + O^{(\ell)}_{00} \mathbb{1}_N \nonumber \\
 &+ \sum_{i,j > 0}  \left(O^{(\ell)}_{ij} - \delta_{i,j}O^{(\ell)}_{00}\right) \sigma_{\ell}^{(i)^{\dagger}} \sigma_{\ell}^{(j)}. \nonumber
\end{align}
Using the inverse of~\eqref{eq:QP_modeop1} and~\eqref{eq:QP_modeop2} we can then use the implicit definition for the \ac{QP} ground state to compute the spectral functions. We note that, while~\eqref{eq:local_OP_representation} is a non-linear representation, \eqref{eq:QP_modeop1} and~\eqref{eq:QP_modeop2} are linear. Thus, due to the implicit definition of the \ac{QP} ground state, only terms of even order in the Gutzwiller operators matter for~\eqref{eq:spec_func} and~\eqref{eq:DSF}. As shown in Sec.~\ref{sec:Commutator_deviation}, the average number $\kappa$ of local Gutzwiller excitations in the \ac{QP} ground state is on the order of a few percent, so we neglect the fourth order terms which would only contribute $\mathcal{O}(\kappa^2)$.

\begin{figure}[t]
 \centering
 \includegraphics[width=0.99\columnwidth]{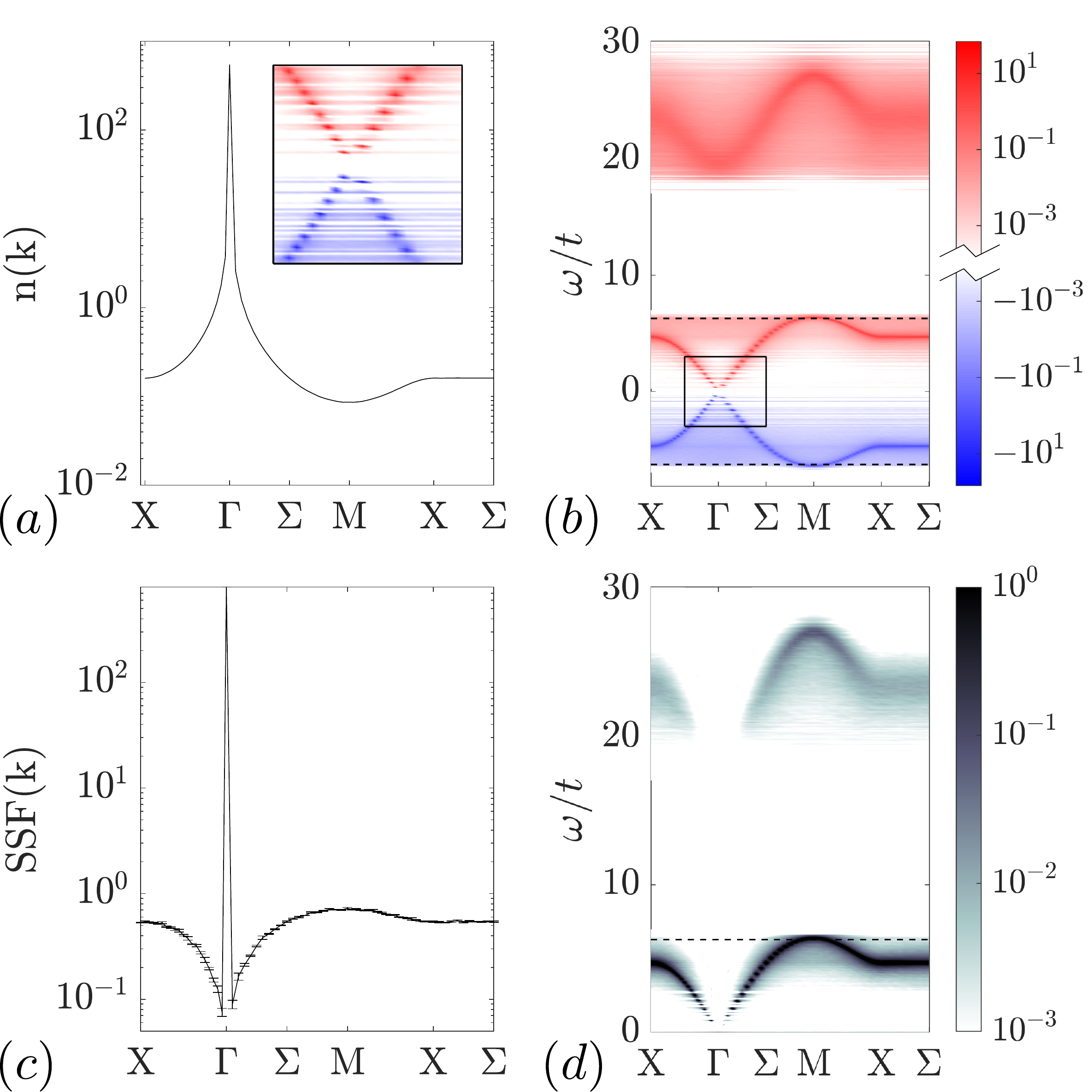}
 \caption{
Finite-size ($L=40$) spectral functions for the QP ground state (see text) of~\eqref{eq:MBL-hamil} at $U/t = 20$ and weak disorder $W/t = 1$. Shown are 
$(a)$ the momentum distribution $n(\mathbf{k})$, [$(b)$, inset $(a)$] the spectral function $\mathcal{A}(\mathbf{k},\omega)$ with separate color-axis for the retarded and advanced branches, $(c)$ the static structure factor $SSF(\mathbf{k})$, and $(d)$ the dynamic structure factor $DSF(\mathbf{k},\omega)$. All spectral functions are shown along a path of high-symmetry points of the first Brillouin-zone of the square-lattice, in units of $\pi/a$: $(0,1) \rightarrow (0,0) \rightarrow (1,1) \rightarrow (1,0) \rightarrow (0.5,0.5)$. Dashed lines in $(b)$ and $(d)$ mark the \ac{ME} as determined in \cite{Geissler2019a}.
}
 \label{fig:QP_spec_W1}
\end{figure}

For $U/t = 20$ we consider a system with 1600 sites ($L=40$) at weak ($W/t = 1$) and moderate ($W/t = 5$) disorder averaged over $N_r = 10$ disorder realizations using a truncation of $N=4$ to discuss signatures of localization in the static properties of the ground state as well as in the spectrum of its \ac{FOE} excitations in relation to the inverse-variance weighted mean of the gap ratio and fractal dimension predictions for the \ac{ME} determined in \cite{Geissler2019a} and shown in Fig.~\ref{fig:amp_decay_U20W10}$(d)$. Firstly, Fig.~\ref{fig:QP_spec_W1} depicts the weak disorder case for which the momentum distribution (panel $a$) has a very pronounced peak at $\mathbf{k} = 0$ corresponding to the condensate fraction while the static structure factor (panel $c$) displays only weak fluctuations due to the disorder but otherwise follows the behavior of the homogeneous case as well. The spectral function (panel $b$, inset panel $a$) and the dynamic structure factor (panel $d$) on the other hand already present strong signatures of localized fluctuations, especially at large \ac{QP} excitation energies in the first gapped band where the spectral weights are spread over all lattice momenta. Conversely, the ungapped (Goldstone) band is well resolved as the \ac{ME} [dashed line in Fig.~\ref{fig:QP_spec_W1}$(b,d)$] is identical to its upper band edge. Especially the low-energy states are almost exactly the low-momentum eigenstates following the linear dispersion of a \ac{SF}.

\begin{figure}[t]
 \centering
 \includegraphics[width=0.99\columnwidth]{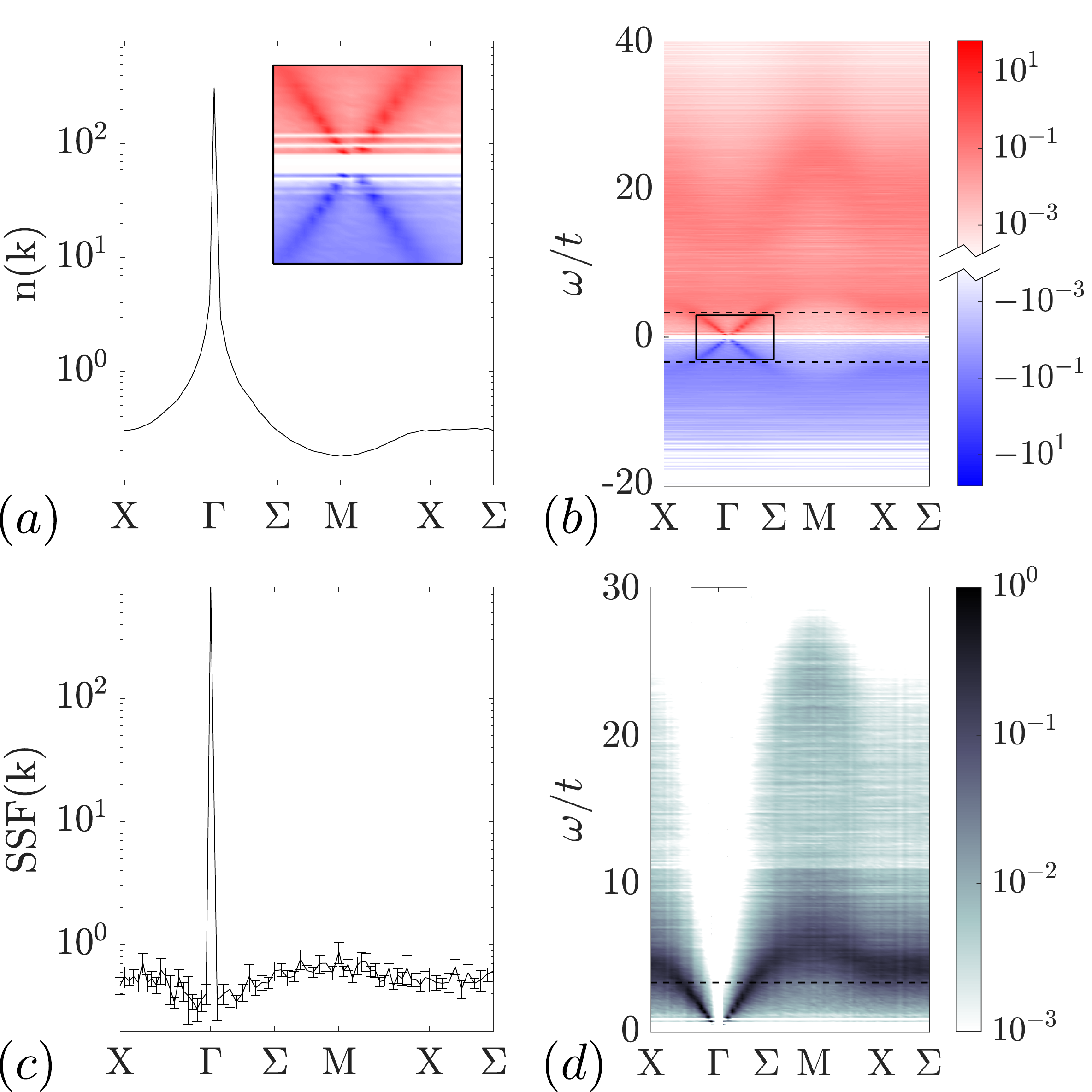}
 \caption{
Finite-size ($L=40$) spectral functions for the QP ground state (see text) of~\eqref{eq:MBL-hamil} at $U/t = 20$ and moderate disorder $W/t = 5$. Shown are $(a)$ the momentum distribution $n(\mathbf{k})$, [$(b)$, inset $(a)$] the spectral function $\mathcal{A}(\mathbf{k},\omega)$ with separate color-axis for the retarded and advanced branches, $(c)$ the static structure factor $SSF(\mathbf{k})$, and $(d)$ the dynamic structure factor $DSF(\mathbf{k},\omega)$. All spectral functions are shown along a path of high-symmetry points of the first Brillouin-zone of the square-lattice, in units of $\pi/a$: $(0,1) \rightarrow (0,0) \rightarrow (1,1) \rightarrow (1,0) \rightarrow (0.5,0.5)$. Dashed lines in $(b)$ and $(d)$ mark the \ac{ME} as determined in \cite{Geissler2019a}.
 }
 \label{fig:QP_spec_W5}
\end{figure}

In contrast, as visible for the spectral function (panel $b$, inset panel $a$) and dynamic structure factor (panel $d$) given in Fig.~\ref{fig:QP_spec_W5}, at an enhanced disorder of $W/t = 5$ the entire spectrum of \ac{QP} states above the \ac{ME} (dashed lines in panels $b$ and $d$) becomes smeared-out over all lattice momenta, while both lowest bands are merging due to the disorder driven local energy fluctuations. Only for \ac{QP} energies below the \ac{ME} one can still find a prevailing linear dispersion of low-momentum \ac{QP} states (inset panel $a$). Regarding the \ac{QP} ground state itself, the increased disorder results in a further decreased zero-momentum peak in $n(\mathbf{k})$ (panel $a$). Also, there is an almost complete loss of non-trivial non-local density correlations, visible in the nearly flat static structure factor (panel $c$) close to the \ac{BG} phase which indicates a nearly uncorrelated distribution of particles. The value of the flat background $s_b = 0.54(3)$ corresponds to the only non-trivial (local) correlations via $c(\mathbf{r}) = N_pL^{-2} \sum_{\mathbf{k}}\exp(i \mathbf{kr})L^{-2} S(\mathbf{k}) \approx \langle \hat{n}_{\ell} \rangle_d (\langle \hat{n}_{\ell} \rangle_d + \delta_{\mathbf{r,0}} s_b)$ where $S(\mathbf{k}) \approx s_b + \delta_{\mathbf{k},0}N_p$. Here, $c(\mathbf{d}) = \sum_{\mathbf{q}} \langle \textrm{QP} | \hat{n}_{\ell} \hat{n}_{\ell'} | \textrm{QP} \rangle_d / L^2$ is the lattice and disorder average of the density correlations where $\mathbf{d} = \mathbf{r}_{\ell} - \mathbf{r}_{\ell'}$ and $\mathbf{q} = \left(\mathbf{r}_{\ell} + \mathbf{r}_{\ell'}\right)/2$. While an uncorrelated placement of particles would imply \ac{P} correlations with $c_{\textrm{P}}(0) = n^2 + n = 3/4$, the local correlations $c(0) = 0.519(14)$ at this disorder are sub-Poissonian due to the repulsive local interactions $U$. This value increases towards the Poissonian value above the critical disorder of the \ac{SF} to \ac{BG} transition. Altogether, this discussion of spectral functions nicely reflects our predictions of the \ac{ME} and is consistent with a superfluid ground state dissolving in favor of a Bose glass phase for increasing disorder. 

\section{Summary}\label{sec:summary}

In this work we have explored the properties of the two-dimensional BHM with disorder, both in the ground state and in its \ac{FOE} quasiparticle spectrum, in order to obtain some insight on the relation between the well-known \ac{BG} ground state phase at moderate disorder and the more elusive localization phenomena of (noninteracting) many-body \ac{QP} excitations at strong disorder. Regarding the \ac{BG} phase, we find that a surprisingly simple fractal dimension analysis of the mean-field ansatz already suffices to reveal a critical disorder strength accompanied by a finite Edwards-Anderson parameter, implying the onset of the \ac{BG} phase. Furthermore, we show that \ac{FOE} gives corrections to this result by considering the finite-size scaling of the lowest energy gaps.

Regarding the \ac{QP} excitations of this corrected ground state we find \ac{QP} level spacing statistics that are consistent with a quasiparticle \ac{ME}. In the localized part of the spectrum fluctuation wave functions have exponential tails which imply exponentially suppressed interactions between localized quasiparticles reminiscent of the local integrals of motion expected for many-body localization. An analysis of the spectral function and dynamic structure factor yields a weak broadening of the spectrum of entangled \ac{QP} excitations below the \ac{ME} while above it they become smeared out over all lattice momenta. In addition, the static structure factor becomes flat at the onset of the \ac{BG} indicating the transition to a phase with vanishing non-local density correlations.

Finally, the \ac{FOE} method arguably yields a very good approximation of the ground state and its \ac{QP} excitations, due to the observed very low fraction of local fluctuations, the interaction of which is neglected when deriving the \ac{FOE} spectrum. As this holds throughout the whole range of considered disorder and local interaction values, we expect the method to be ideally suited to evaluate the dynamics of typical experimental quenching protocols, for example in order to determine the evolution of the entanglement entropy after a sudden quench of the disorder potential. Furthermore, considering their interactions the \ac{QP} modes yield a promising basis for the construction of \ac{LIOM}s and to study the stability of the non-interacting \ac{ME}. \newline

\begin{acknowledgments}
The author would like to thank L. Rademaker for insightful discussions and especially G. Pupillo for his extensive support and many comments. Support by the Leopoldina Fellowship Programme of the German National Academy of Sciences Leopoldina grant no. LPDS 2018-14, the ANR ERA-NET QuantERA - Projet RouTe (ANR-18-QUAN-0005-01) and the High Performance Computing center of the University of Strasbourg, providing access to computing resources and scientific support, is gratefully acknowledged. Part of the computing resources were funded by the Equipex Equip@Meso project (Programme Investissements d'Avenir) and the CPER Alsacalcul/Big Data.
\end{acknowledgments}

\bibliography{MBLlib}

\end{document}